\documentclass[aps,eqsecnum,showpacs]{revtex4}
\usepackage{epsfig}
\usepackage{amsmath}

\begin{document}

%\begin{flushright}
%\today  \end{flushright}

\title{
The n+n+alpha System in a Continuum Faddeev Formulation
}

\author{K. Khaldi$^{1,2}$}

\author{Ch. Elster$^1$}

\author{W.~Gl\"ockle$^3$}

\affiliation{$^1$Institute of Nuclear and Particle Physics,
Department of Physics and Astronomy, Ohio University, Athens, OH 45701, USA}

\affiliation{$^2$ Facult\'e des Sciences, Universit\'e de Boumerdes, 35000 Boumerdes,
Algeria}

\affiliation{$^3$Institut f\"ur Theoretische Physik II,
Ruhr-Universit\"at Bochum, D-44780 Bochum, Germany}

\vspace*{10mm}

\date{\today}

\vspace*{10mm}

\begin{abstract}

The continuum Faddeev equations for the neutron-neutron-alpha (n+n+$\alpha$)
 system are formulated for a
general interaction as well as for finite rank forces. In addition, the capture process 
n+n+$\alpha \rightarrow ^6$He+$\gamma$ is derived. 

\end{abstract}


\maketitle

\setcounter{page}{1}

\section{Introduction}

In recent years the study of quantum halo systems experienced increased
interest in the nuclear as well as the atomic few-body community. For a recent
review see Ref.~\cite{Jensen:2004zz}. The nucleus $^6$He is of particular
interest, since it constitutes the lightest two-neutron halo nucleus with
a $^4$He core. Being an effective three-body system, the properties of the
ground state have been explored using either 
Faddeev~\cite{lehman1,lehman2,lehman3} 
or Hyper-spherical Harmonics (HH)~\cite{danilin1,Zhukov:1993aw,danilin3,Fedorov:2003jx}. 
More recently
the ground state has also been calculated with multi-cluster
methods~~\cite{Csoto:1993fg,Varga:1994fu,kukulin,aoyama1,aoyama2,Wurzer:1996rr,
suzuki2,Kato,Hiyama} as well
as with  GFMC~\cite{Pudliner:1995wk}.
Those multi-cluster methods include 
various techniques like 
the microscopic dynamical multi-configuration three-cluster model~\cite{Csoto:1993fg}, the
stochastic variational method~\cite{Varga:1994fu}, the multi-cluster dynamic model (MDMP
and AMDMP), the hybrid-TV model, 
a combination of the cluster orbital shell
model (COSM)~\cite{suzuki2} and the extended cluster model (ECM)~\cite{Kato}, 
the refined resonating group method~\cite{kukulin,Wurzer:1996rr}, and the
coupled-rearrangement-channel variational method with  Gaussian  basis
functions~\cite{Hiyama}.
In addition, the beta decay to the alpha+d continuum has been 
studied~\cite{Zhukov:1992az,Tursunov:2005dv}.
 Of interest is also the two-neutron capture process $^4$He(2n,$\gamma$)$^6$He 
as a possible route bridging the instability gap at nuclear mass A=5~\cite{efros}.

The situation is quite different in the continuum of two neutrons and an
$\alpha$ -particle. There is a  well established $2^+$  resonance
\cite{ajzenberg}, but further resonant structures are still under debate
\cite{ershov1,Danilin:1997zz,ershov2,Danilin:2004mx,
danilin3,Danilin:2006qq,ershov3,Pieper:2004qw,Descouvemont:2005rc}. 
Up to now more indirect approaches in understanding the resonance structure have been
carried out, e.g. the four-body distorted wave approach, leading to three-body continuum
excitations of two-neutron Borromean halo
nuclei~\cite{ershov1,ershov3}. Furthermore, complex scaling in Coulomb break-up reactions has been
employed~\cite{Kikuchi:2010zz}.
In addition, an extension of the HH method on a Lagrange mesh~\cite{Descouvemont:2005rc}
has been used to study three-body continuum states.  This is at least a four-body problem
with great uncertainties about the reaction
 mechanisms and the interactions entering these much more complicated systems.

Thus, the currently predominant approach to continuum
calculations for the pure $n+n+\alpha$ system is the HH
method~\cite{danilin3,Fedorov:2003jx,Danilin:2004mx,Danilin:2006qq,Danilin:2007zz,ershov3}.
A Faddeev approach is to the best of our knowledge still missing. Only for
the $^6$Li nucleus, a Faddeev treatment of the deuteron-alpha (d-$\alpha$) system has been
employed~\cite{Eskandarian:1992zz}, which however, did not have to face the
challenge of  three-to-three scattering. This also refers to the pioneering work 
by Koike~\cite{Koike1,Koike2} on d-$\alpha$ scattering.

The  aim of this investigation  is to fill that gap. For the $n+n+\alpha$
system one faces the situation of three free particles being in the
initial channel and leading again to three free particles in the exit
channel. In other words, one has to deal with three-to-three scattering.  
Scattering of three free incoming particles to three free outgoing ones in a Faddeev
approach has been initiated in~\cite{barbour,gibson} in the context of the three-body photo
disintegration of $^3$He. This path has also been followed in the same context and in a
Faddeev approach by Meijgaard and Tjon~\cite{vanMeijgaard:1992zz}.
Into the matrix element for the photo disintegration enters 
the three-nucleon to three-nucleon
scattering wave function, which has been evaluated in Ref.~\cite{vanMeijgaard:1992zz} and then
inserted into the photodisintegration matrix element.
 However, evaluating the wave function is
  a completely unnecessary complication, since this process is
initiated by the three-nucleon bound state. One can  directly derive
  a Faddeev equation for the three-body break up amplitude, in which  the driving term
contains the action of the current  operator on the $^3$He ground state.
Then, the complete final state interaction  is generated by a Faddeev integral kernel
for the amplitude. This considerably simplifies the technical part of a calculation, 
since no disconnected processes occur. 
This very procedure has been pioneered in Refs.~\cite{barbour,gibson} and is being
applied in state-of-the-art calculations, see e.g. \cite {Golak:2005iy, wgphysrep}.

The same procedure  can trivially be adapted to the capture process $ n+n+\alpha
\rightarrow ^6$He$+\gamma$, as  will be displayed in the present investigation. 
   This capture process is relevant for the production rate
of $^6$He  in astrophysical environments~\cite{deDiego:2010sf}
characterized by high neutron and alpha densities e.g. those related to supernova shock
fronts. In e.g. Ref.~\cite{efros} this three-body process is approximated by sequential
two-body processes, whereas in principle a genuine three-body reactions needs to
be calculated.  Furthermore, the $nn\alpha \rightarrow nn\alpha$ amplitude is relevant for
determining the the next order coefficient~\cite{Dashen}
in the virial equation of state in low-density
matter~\cite{Horowitz:2005nd}.

From a technical point of view the Faddeev approach to the $ n+n+\alpha$ continuum is
strongly needed, since the
 currently predominant approach, namely the Hyper-spherical Harmonics (HH) approach, still
faces  open challenges.
It is already known that in the break up process  $n+d \rightarrow n+n+p$
  a strong FSI peak appears  for the $n-n$ subsystem.
In the Faddeev approach using Jacobi momenta this can be mapped out correctly,
whereas when changing to the hyper-spherical angle, the convergences is quite poor for this
particular configuration.
In the HH method, the control of the coupling potentials
can be a painful exercise, whereas in the Faddeev approach using Jacobi variables
the dynamics is perfectly well under control in all details.
This same situation must be expected in $n+n+\alpha$ scattering, where the three-body
S-matrix is characterized by continuous quantum numbers describing how the energy is
distributed over the relative motion. There are strong initial and final state
interaction peaks, which in a discrete representation through hyper-spherical
K-quantum numbers are  difficult to map out correctly.
As stated above only a technically reliable approach as the Faddeev one will guarantee the
validity of the results when searching for $^6$He resonances.

The paper is organized as follows. In Section II we derive the coupled Faddeev
equations for the three-to-three scattering amplitude, followed by a partial wave
decomposition in Section III. 
The Faddeev equations will be solved by iteration yielding a multiple
scattering series. This will be outlined in Section IV. 
Since most of the Faddeev type investigations of the $n+n+\alpha$ system
are based on finite rank forces, we also present in Section V a continuum 
formulation based on separable forces. Furthermore, we discuss the 
the unitarity relation for the three-to-three amplitude in Section VI. 
Finally the capture process  $n+n+\alpha \rightarrow ^6$He+$\gamma$ 
will be discussed for the Faddeev scheme in Section VII. 
Then we summarize in section VII. 
Technical details about the partial wave decomposition and an efficient
way of treating the three-body singularities are given in the Appendices.

\section{The Faddeev Equations for the nn$\alpha$ System}
\label{section2}

In developing the formal expression for the transition amplitude between
three free particles
interacting with short-range, strong interactions, we start from the triad
of Lippmann-Schwinger (LS)
equations~\cite{wg1970,wbook} acting on a three-particle initial
state given by
\begin{equation}
\Phi_{\alpha}^{(+)} =  | {\bf p}_{\alpha}\rangle^{(+)} | {\bf
q}_{\alpha}
\rangle,
\label{eq:1.1}
\end{equation}
where $|{\bf p}_{\alpha} \rangle ^{(+)}$ is a two-body scattering state,
and the index
$\alpha = 1,2,3$ indicates the three choices of pairs characterized by the third
particle, the spectator. 
Furthermore,  $ V^{ \alpha} = \sum_{ \beta \ne \alpha} V_{\beta} $ , where $ V_{\beta}$
($\beta = 1,2,3$)  are the pair forces. Three-body forces can in principle be
incorporated in a straightforward fashion. However, we will only concentrate on two-body
forces here. The triad of LS equations,
\begin{eqnarray}
\Psi_0^{(+)} = \Phi_{\alpha}^{(+)} + G_{\alpha} V^{\alpha} \Psi_0^{(+)},
\label{eq:1.2}
\end{eqnarray}
define the scattering wave uniquely.
The channel  Green's function is given by $G_{\alpha}^{-1}=(E+i\varepsilon -H_0
-V_{\alpha})^{-1}$.
We use standard Jacobi momenta ${\bf p}_{\alpha}$ and ${\bf q}_{\alpha}$ and their
quantum numbers as basis
states.

By suitable multiplication of the three
 equations in the triad from the left by $ V_{\beta} $  one obtains the transition
operators
$ U_{ \alpha 0} \equiv ( V_{\beta} + V_{\gamma} ) \Psi_0^{(+)}$, with
$\beta \ne \alpha, \gamma \ne \alpha $, which fulfill the set of equations
\begin{eqnarray}
U_{ \alpha 0} = \sum_{ \beta \ne \alpha} t_{\beta} \Phi_0 + \sum_{ \beta \ne \alpha} 
t_{\beta} G_0 U_{\beta 0},
\label{eq:1.3}
\end{eqnarray}
where $\Phi_0=|{\bf p}\rangle |{\bf q}\rangle$ is the free three-particle state.

The three-body break up operator is given by
\begin{eqnarray}
U_{00} \equiv \sum_{\gamma} V_{\gamma} \Psi_0^{(+)}.
\label{eq:1.4}
\end{eqnarray}
Again, from the triad follows
\begin{eqnarray}
U_{00} =\sum_{\gamma} t_{\gamma}  \Phi_0 + \sum_{\gamma} t_{\gamma} G_0 U_{\gamma 0}.
\label{eq:1.5}
\end{eqnarray}

\noindent
Iterating Eq.~(\ref{eq:1.3}) one obtains the multiple scattering series
\begin{eqnarray}
U_{00} =\sum_{\gamma} t_{\gamma}  \Phi_0  + \sum_{\gamma} t_{\gamma}G_0 
\sum_{\beta \ne \gamma} t_{\beta} \Phi_0
+ \sum_{\gamma} t_{\gamma} G_0\sum_{\beta \ne \gamma} t_{\beta} G_0 
\sum_{ \alpha \ne \beta} t_{\alpha} \Phi_0 + \cdots \; .
\label{eq:1.6}
\end{eqnarray}

Instead of working with the coupled set of  Eq.~(\ref{eq:1.3})  and the relation 
of  Eq.~(\ref{eq:1.5})  for the three-body break-up operator, one can generate the
multiple scattering series directly by
 decomposing $ U_{00} $ as

\begin{eqnarray}
U_{00} \equiv \sum_{\gamma} U_{\gamma},
\label{eq:1.7}
\end{eqnarray}
and choosing $ U_{\gamma} $ to obey the coupled set of Faddeev equations
\begin{eqnarray}
U_{\gamma} = t_{\gamma} + t_{\gamma} G_0 \sum_{ \alpha \ne \gamma} U_{ \alpha}.
\label{eq:1.8}
\end{eqnarray}
Indeed, iterating Eq.~(\ref{eq:1.8}) and inserting the result into Eq.~(\ref{eq:1.7})
leads exactly to the multiple scattering series from above.
Explicitly, we have a set of three coupled equations
\begin{eqnarray}
U_1 &=& t_1 + t_1 G_0 ( U_2 + U_3)\cr
U_2 &=& t_2 + t_2 G_0 ( U_3 + U_1)\cr
U_3 &=& t_3 + t_3 G_0 ( U_1  + U_2).
\label{eq:1.9}
\end{eqnarray}
We also observe, that comparing Eqs.~(\ref{eq:1.7}) and (\ref{eq:1.4}) leads to
\begin{eqnarray}
U_{\gamma} \equiv V_{\gamma} \Psi_0^{(+)}.
\label{eq:1.10}
\end{eqnarray} 

When considering the n+n+$\alpha$ system, we need to incorporate the identity of the
two neutrons. Fixing arbitrarily the alpha particle as spectator and label it as ``1''
and the two neutrons as particles ``2'' and ``3'', the scattering wave function
$\Psi_0^{(+)}$ must be antisymmetric under the exchange of particles ``2'' and ``3''.
Thus, defining the transposition operator $P_{23}$, the scattering wave function must
fulfill $ P_{23}\Psi_0^{(+)} = - \Psi_0^{(+)}$. Using this in Eq.~(\ref{eq:1.10}) leads
to 
\begin{equation}
U_3 = - P_{23} U_2.
\label{eq:1.11}
\end{equation}
Thus, for the n+n+$\alpha$ system we only have 2 coupled equations,
\begin{eqnarray}
U_1 &=& t_1 + t_1 G_0 \; ( 1 - P_{23}) \;  U_2 \cr
U_2  &=& t_2 + t_2 G_0 \; (-P_{23}  U_2 + U_1)
\label{eq:1.12}
\end{eqnarray}
More precisely, one has to apply the driving terms to the free state $\Phi_{0,a}$, 
which is antisymmetric under exchange of the two neutrons:
\begin{equation}
\Phi_{0,a} \equiv ( 1 - P_{23}) | {\bf p}_1 {\bf q}_1 \rangle | 0 m_2 m_3 \rangle 
\Big| 0 \frac{1}{2}\frac{1}{2} \Big\rangle,
\label{eq:1.13}
\end{equation}
leading to
\begin{eqnarray}
U_1\Phi_{0,a}  &=& t_1\Phi_{0,a} + t_1 G_0  \left( 1   -P_{23}\right)   U_2 \Phi_{0,a}\cr
U_2\Phi_{0,a}  &=& t_2\Phi_{0,a} + t_2 G_0  \left(-  P_{23}  U_2\Phi_{0,a} +
U_1\Phi_{0,a}\right).
\label{eq:1.14}
\end{eqnarray}
The full break-up operator is given by
\begin{eqnarray}
U_{00}\Phi_{0,a}  = U_1\Phi_{0,a} + ( 1 - P_{23}) U_2\Phi_{0,a}.
\label{eq:1.15}
\end{eqnarray}
For the on shell break-up amplitude one has to evaluate the matrix element
$\langle \Phi_{0,a}' |U_{00,a}| \Phi_{0,a}\rangle $,
where in the final state momenta as well as spin magnetic quantum numbers are changed,
\begin{eqnarray}
\Phi_{0,a}' = ( 1 - P_{23}) | {\bf p}_1' {\bf q}_1' \rangle | 0 m_2' m_3'\rangle \Big| 0
\frac{1}{2}\frac{1}{2} \Big\rangle.
\label{eq:1.16}
\end{eqnarray}

\section{Partial Wave Decomposition}
\label{section3}

In order to solve the coupled equations, Eqs.~(\ref{eq:1.14}), two sets of 
partial wave basis states are needed:  
\begin{eqnarray}
| p_1 q_1 \alpha_1\rangle & \equiv &  \sum_{\mu_1} C(j_1 \lambda_1 J, \mu_1 M_1 - \mu_1) | 
p_1 ( l_1 s_1 ) j_1 \mu_1\rangle | q_1 \lambda_1 M_1 - \mu_1\rangle \Big| \left( \frac{1}{2}
\frac{1}{2} \right) 1 \Big \rangle \cr
| p_2 q_2 \alpha_2\rangle & \equiv&  \sum_{\mu_2} C( j_2 I_2 J, \mu_2 M_2 - \mu_2) | p_2
( l_2 s_2 ) j_2 \mu_2\rangle | q_2 (\lambda_2 \frac{1}{2}) I_2  M_2 - \mu_2\rangle \cr 
& & \Big| \left( \frac{1}{2} \frac{1}{2}\right) 1 \Big\rangle \; .
\label{eq:1.17}
\end{eqnarray}
The details of a partial wave decomposition of Eqs.~(\ref{eq:1.14})  is well known 
(see e.g. \cite{wgphysrep}),  and we refer to \cite{khalida}  for details. 
Employing the states of Eq.~(\ref{eq:1.17}), the coupled equations, Eq.~(\ref{eq:1.14})
read  
 \begin{eqnarray}
\lefteqn{ \langle p_1' q_1' \alpha_1'| U_{1,a}  = } \cr
& &   \frac{\delta( q_1' - q_1)}{ q_1^2} \; t_{\alpha_1'}( p_1' p_1, E_{q_1}) \;
 C_{\alpha_1'}^{ m_2 + m_3} ( \theta_1) \cr
  & + &  \left( 1 + ( -1)^{ l_1' + s_1'}\right) \int dx \int dq_2' q_2^{'2} \;
  t_{ \alpha_1'} ( p_1' \pi_1 ( q_1' q_2' x), E_{q_1'}) \; G_0(  \pi_1 ( q_1' q_2' x), q_1')\cr
  & & \sum_{\alpha_2'} G_{ \alpha_1' \alpha_2'} ( q_1' q_2' x) \langle \pi_2 ( q_1' q_2' x) q_2' \alpha_2'| U_{2,a}\cr
\lefteqn{ \langle p_2' q_2' \alpha_2'| U_{2,a} =} \cr
& &  \frac{ \delta( q_2' - q_2)}{ q_2^2} \; t_{ \alpha_2'} ( p_2' p_2, E_{q_2}) \;
 D_{ \alpha_2'}^{ m_2, m_3} ( \theta_1)\cr
  & - & \frac{ \delta( q_2' - \tilde q_2)}{ {\tilde q_2}^2} \; 
t_{ \alpha_2'} ( p_2' \tilde p_2, E_{ \tilde q_2}) \;
  \tilde D_{ \alpha_2'}^{ m_2, m_3} ( \theta_1) \cr
  & + &  \int dx  \int  dq_1' q_1^{'2} \; t_{\alpha_2'} ( p_2' \pi_3 ( q_2' q_1' x),
E_{q_2'}) \; G_0( \pi_3 ( q_2' q_1' x),q_2')\cr
  & & \sum_{\alpha_1'} H_{ \alpha_2' \alpha_1'}( q_2' q_1' x) \langle \pi_4 ( q_2' q_1' x) q_1' \alpha_1' | U_{1,a}\cr
  & - &    \int dx \int dq_2^{'''} q_2^{'''2} \; t_{\alpha_2'} ( p_2' \pi_5 ( q_2'
q_2^{'''} x), E_{q_2'}) \;  G_0( \pi_5 ( q_2' q_2^{'''} x),q_2')\cr
  & & \sum_{\alpha_2^{'''}} I_{ \alpha_2' \alpha_2^{'''}} ( q_2' q_2^{'''} x) \langle
\pi_6( q_2' q_2^{'''} x) q_2^{'''} \alpha_2^{'''} | U_{2,a} \; ,
\label{eq:1.18}
  \end{eqnarray}
where
\begin{eqnarray}
C_{\alpha_1}^{ m_2 + m_3} ( \theta_1) & = &  \left( 1 + (-)^{ l_1 + s_1}\right) 
\left( \frac{1}{2}\frac{1}{2} s_1, m_2 m_3 \right) \sum_{ m_{l_1}} ( l_1 s_1 j_1, m_{l_1}, m_2 + m_3)\cr
& & ( j_1 \lambda_1 J, m_{l_1} + m_2 + m_3, 0 , M) Y_{ l_1 m_{l_1}} ( \theta_1,0) \sqrt{
\frac{\hat \lambda_1}{ 4 \pi}} \; ,
\label{eq:1.19}
\end{eqnarray}
with $\hat \lambda_1 = 2 \lambda_1 +1$.
\begin{eqnarray}
\lefteqn{ D_{ \alpha_2'}^{ m_2, m_3} ( \theta_1) \equiv D_{ \alpha_2'}^{ m_2, m_3} (
\theta_{p_2} \theta_{q_2}) }\cr
&  = &  \delta_{ s_2' \frac{1}{2}} \sum_{\mu} \left( j_2' I_2' J', \mu,
M'-\mu\right)\left( l_2' \frac{1}{2}  j_2', \mu - m_3, m_3\right) \;  
Y_{ l_2' \mu - m_3}^* ( \hat p_2)\cr
& & \left( \lambda_2'  \frac{1}{2} I_2', M' -\mu - m_2, m_2\right) 
Y_{ \lambda_2' M' -\mu - m_2}^* ( \hat q_2)
\label{eq:1.20}
\end{eqnarray}
and
\begin{eqnarray}
\lefteqn{\tilde D_{ \alpha_2'}^{ m_2, m_3} ( \theta_1) \equiv \tilde D_{ \alpha_2'}^{
m_2, m_3} ( \theta_{p_2} \theta_{q_2}) } \cr
& = &  \delta_{ s_2' \frac{1}{2}} \sum_{\mu} \left( j_2' I_2' J', \mu
M'-\mu\right)\left( l_2'\frac{1}{2} j_2', \mu - m_2, m_2\right) \; 
Y_{ l_2' \mu - m_2}^* ({\hat {\tilde p}}_2)\cr
& & \left( \lambda_2'  \frac{1}{2} I_2', M' -\mu - m_3, m_3\right) \; 
Y_{ \lambda_2'M' -\mu - m_3}^* ({\hat {\tilde q}}_2) \; .
\label{eq:1.21}
\end{eqnarray}
The `shifted' momenta $\pi_i$ are given as
\begin{eqnarray}
\pi_1 &=& \sqrt{ \alpha^2 q_1^{'2} + q_2^{'2} + 2 \alpha q_1' q_2' x} \cr
\pi_2 &=& \sqrt{  q_1^{'2} + \beta^2 q_2^{'2} + 2 \beta q_1' q_2' x} \cr
\pi_3 &=& \sqrt{  q_2^{'2} + \beta^2 q_1^{'2} + 2 \beta q_2' q_1' x} \cr
\pi_4 &=& \sqrt{  \alpha^2 q_2^{'2} +  q_1^{'2} + 2 \alpha q_2' q_1' x} \cr
\pi_5 &=& \sqrt{  {\overline \beta}^2 q_2^{'2} +  q_2^{'''2} + 2 {\overline \beta} q_2' q_2^{'''} x}\cr
\pi_6 &=& \sqrt{ q_2^{'2} +  {\overline \beta}^2 q_2^{'''2} + 2 {\overline \beta} q_2'
q_2{'''} x} \; ,
\label{eq:1.22}
\end{eqnarray}
where
\begin{eqnarray}
\alpha & = & \frac{1}{2}\cr
\beta & = &  \frac{ m_{\alpha}}{ m + m_{\alpha}}\cr
\gamma & = &  \frac{ 2 m + m_{\alpha}}{ 2( m + m_{\alpha})}\cr
\overline \beta & = &  \frac{ m}{ m + m_{\alpha}} \; .
\label{eq:1.23}
\end{eqnarray}
Here $m$ is the neutron mass, and $m_\alpha$ the mass of the $^4$He nucleus. 

We refer to the Appendix~\ref{appendixa}  
for some details of the derivation and the expressions of 
the purely geometric quantities $G_{ \alpha_1' \alpha_2'} ( q_1' q_2' x)$, 
$H_{ \alpha_2' \alpha_1'}( q_2' q_1' x)$, and  $I_{ \alpha_2' \alpha_2^{'''}} ( q_2'
q_2^{'''} x)$. Furthermore,  $ t_{\alpha_1'}( p_1' p_1, E_{q_1})$ is the two-neutron
$t$-matrix and 
$t_{ \alpha_2'} ( p_2' p_2, E_{ q_2})$ the one for the neutron-$\alpha$ pair.

Due to the  free Green's functions $G_0$  and the x- integration over it,
 one encounters  the
well known  logarithmic singularities of any three-body problem. These singularities
 can be  reliably treated \cite{wgphysrep,Liu:2004tv}. However, the method
suggested in Refs.~\cite{Witala:2008my,Elster:2008hn},  appears to be beneficial for
here,
since not only kernels contain logarithmic singularities but also the driving
terms. We illustrate this new method with an example in Appendix~\ref{appendixb}.

\section{The Multiple Scattering Series}
\label{section4}

\begin{figure}[b]
\begin{center}
 \includegraphics[width=13cm]{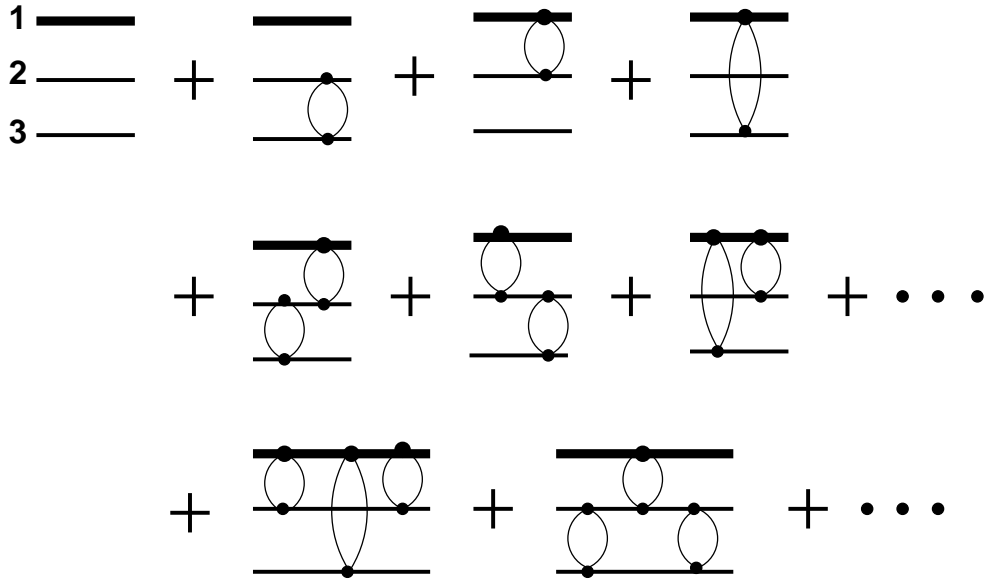}
\end{center}
\caption{Diagrammatic representation of the first few terms of the
multiple scattering series for the neutron-neutron-$\alpha$ system.
Here the alpha particle ($1$) is indicated by the thicker line.
\label{fig1}}
\end{figure}

A  well established way to solve a coupled set of Faddeev equations
is to generate the multiple scattering series.
For the 3N system is is laid  out e.g. in Ref.~\cite{wgphysrep}.
Schematically the Eqs.~(\ref{eq:1.18}) have the form
\begin{eqnarray}
  U = U^{(0)} + K U
\label{eq:1.24}
\end{eqnarray}
 which,  when iterated yield
  \begin{eqnarray}
  U = U^{(0)} + U^{(1)} + U^{(2)} + \cdots  \; ,
\label{eq:1.25}
  \end{eqnarray}
   with
  \begin{eqnarray}
  U^{(n)} = K U^{(n-1)}, n=1,2,\cdots \; .
\label{eq:1.26}
  \end{eqnarray}
The first few terms of this series are depicted in Fig.~\ref{fig1}
The driving terms of Eqs.~(\ref{eq:1.18}), sketched in the upper row of  Fig.~\ref{fig1}
  are necessarily disconnected, since a two-body t-matrix can not act on three
particles.

  Let us consider the terms of second order in the two-body t-matrix (indicated in the
second row of  Fig.~\ref{fig1}):
  \begin{eqnarray}
\lefteqn{ \langle p_1' q_1' \alpha_1'| U_{1,a}^{(1)}  \equiv } \cr
& & \left( 1 + ( -1)^{ l_1' + s_1'}\right) \int dx \int dq_2' q_2^{'2} \;
   t_{ \alpha_1'} ( p_1' \pi_1 ( q_1' q_2' x), E_{q_1'}) \;
    G_0(  \pi_1 ( q_1' q_2' x), q_1') \cr
& & \sum_{\alpha_2'} G_{ \alpha_1' \alpha_2'} ( q_1' q_2' x)
 \Bigg[  \frac{ \delta( q_2' - q_2)}{ q_2^2} \;  t_{ \alpha_2'} ( \pi_2 ( q_1' q_2' x)
p_2, E_{q_2}) \;  D_{ \alpha_2'}^{ m_2, m_3} ( \theta_1) \cr
& - & \frac{ \delta( q_2' - \tilde q_2)}{ {\tilde q_2}^2} \; 
t_{ \alpha_2'} ( \pi_2 ( q_1' \tilde q_2 x) \tilde p_2, E_{\tilde q_2}) \;
  \tilde D_{ \alpha_2'}^{ m_2, m_3} ( \theta_1) \Bigg]\cr
  & = & \left( 1 + ( -1)^{ l_1' + s_1'} \right) \int dx \sum_{\alpha_2'}
 \; \Bigg[ t_{ \alpha_1'} ( p_1' \pi_1 ( q_1' q_2 x), E_{q_1'})\;  
G_0(  \pi_1 ( q_1' q_2 x), q_1') \cr
& & G_{ \alpha_1' \alpha_2'} ( q_1' q_2 x) \;
    t_{ \alpha_2'} ( \pi_2 ( q_1' q_2 x) p_2, E_{ q_2}) \;  
D_{ \alpha_2'}^{ m_2, m_3} ( \theta_1)\cr
& - &  t_{ \alpha_1'} ( p_1' \pi_1 ( q_1' \tilde q_2 x), E_{q_1'}) \; 
G_0(  \pi_1 ( q_1' \tilde q_2 x), q_1') \;  G_{ \alpha_1' \alpha_2'} ( q_1' \tilde q_2 x) \cr
& & t_{ \alpha_2'} ( \pi_2 ( q_1' \tilde q_2 x) \tilde p_2, E_{\tilde
q_2} ) \; \tilde D_{ \alpha_2'}^{ m_2, m_3} ( \theta_1) \Bigg] \; .
\label{eq:1.27}
  \end{eqnarray}
The only singular function under the x-integral is the free Green's function,  which
 leads in the $ q_1'-q_2$ and $ q_1' - \tilde q_2$  planes of external momenta 
to the well known logarithmic singularities.
The same is true for
\begin{eqnarray}
\lefteqn{ \langle p_2' q_2' \alpha_2'| U_{2,a}^{(1)} \equiv } \cr
& &  \int dx  \int  dq_1' q_1^{'2} t_{\alpha_2'} ( p_2' \pi_3 ( q_2' q_1' x), E_{ q_2'})
\; G_0( \pi_3 ( q_2' q_1' x),q_2') \cr
  & & \sum_{\alpha_1'} H_{ \alpha_2' \alpha_1'}( q_2' q_1' x) \; 
\frac{\delta(q_1'-q_1)}{ q_1^2} \;  t_{\alpha_1'}( \pi_4 ( q_2' q_1' x) p_1, E_{q_1}) \;
  C_{\alpha_1'}^{ m_2 + m_3} ( \theta_1)\cr
 &-& \int dx \int dq_2^{'''} q_2^{'''2} \; t_{\alpha_2'} ( p_2' \pi_5 ( q_2' q_2^{'''}
x),E_{ q_2'}) \; G_0( \pi_5 ( q_2' q_2^{'''} x),q_2') \cr
& &  \sum_{\alpha_2^{'''}} I_{ \alpha_2' \alpha_2^{'''}} ( q_2' q_2^{'''} x)
    \Bigg[ \frac{ \delta( q_2^{'''} - q_2)}{ q_2^2} \; 
t_{ \alpha_2^{'''}} (\pi_6( q_2' q_2^{'''} x) p_2, E_{q_2}) \;
   D_{ \alpha_2^{'''}}^{ m_2, m_3} ( \theta_1) \cr
   & - & \frac{ \delta( q_2^{'''} - \tilde q_2)}{ {\tilde q_2}^2}  \; 
t_{ \alpha_2^{'''}} ( \pi_6( q_2' q_2^{'''} x) \tilde p_2, E_{ \tilde q_2}) \;
   \tilde D_{ \alpha_2^{'''}}^{ m_2, m_3} ( \theta_1) \Bigg]\cr
&=& \int dx \;  t_{\alpha_2'} ( p_2' \pi_3 ( q_2' q_1 x), E_{q_2'}) \; 
G_0( \pi_3 ( q_2' q_1 x),q_2') \cr
   & & \sum_{\alpha_1'}  H_{ \alpha_2' \alpha_1'}( q_2' q_1 x) \; 
t_{\alpha_1'}( \pi_4 ( q_2' q_1 x) p_1, E_{q_1}) \; C_{\alpha_1'}^{ m_2 + m_3} ( \theta_1)\cr
&-&  \int dx \; \Bigg[ t_{\alpha_2'} ( p_2' \pi_5 ( q_2' q_2 x), E_{q_2'}) \;
    G_0( \pi_5 ( q_2' q_2 x),q_2') \cr
& & \sum_{\alpha_2^{'''}} I_{ \alpha_2' \alpha_2^{'''}} ( q_2' q_2 x) \;
 t_{ \alpha_2^{'''}} ( \pi_6( q_2' q_2 x) p_2, E_{q_2}) \;
   D_{ \alpha_2^{'''}}^{ m_2, m_3} ( \theta_1) \cr
& - & t_{\alpha_2'} ( p_2' \pi_5 ( q_2' \tilde q_2 x), E_{q_2'}) \;
    G_0( \pi_5 ( q_2' \tilde q_2 x),q_2') \cr
 & & \sum_{\alpha_2^{'''}} I_{ \alpha_2' \alpha_2^{'''}} ( q_2' \tilde q_2 x) \;
 t_{ \alpha_2^{'''}} ( \pi_6( q_2' \tilde q_2 x) \tilde p_2, E_{\tilde q_2}) \;
   \tilde D_{ \alpha_2^{'''}}^{ m_2, m_3} ( \theta_1) \Bigg] \; .
\label{eq:1.28}
  \end{eqnarray}
Indeed, the free Green's functions $ G_0$ lead in the only remaining  x-integral to 
logarithmic singularities.

In order to safely apply the kernel to the previous amplitude, that amplitude has to be 
a smooth function. This is only the case for the
   next higher order, being of third order in $t$, sketched in the third row of 
Fig.~\ref{fig1}. The third order term reads  
\begin{eqnarray}
\lefteqn{ \langle p_1' q_1' \alpha_1'| U_{1,a}^{(2)} \equiv }\cr
& & \left( 1 + ( -1)^{ l_1' + s_1'}\right) \int dx \int dq_2' q_2^{'2} \;
   t_{ \alpha_1'} ( p_1' \pi_1 ( q_1' q_2' x), E_{q_1'})  \;
 G_0(  \pi_1 ( q_1' q_2' x), q_1') \cr
& &\sum_{\alpha_2'} \; G_{ \alpha_1' \alpha_2'} ( q_1' q_2' x) \;
 \Bigg[  \int dy \;  t_{\alpha_2'} ( \pi_2( q_1' q_2' x)  \pi_3 ( q_2' q_1 y ),
E_{q_2'}) \; G_0( \pi_3 ( q_2' q_1 y ),q_2')\cr
 & & \sum_{\alpha_1^{''}} H_{ \alpha_2' \alpha_1^{''}}( q_2' q_1 y ) \; 
t_{\alpha_1^{''}}( \pi_4(q_2' q_1 y ) p_1, E_{q_1}) \; C_{\alpha_1^{''}}^{m_2 +m_3}(\theta_1)\cr
 & - & \int dy \;  \Bigg[ t_{\alpha_2'} ( \pi_2( q_1' q_2' x) \pi_5 ( q_2' q_2 y ),
E_{q_2'})\; G_0( \pi_5 ( q_2' q_2 y ),q_2') \cr
    & & \sum_{\alpha_2^{'''}} I_{ \alpha_2' \alpha_2^{'''}} ( q_2' q_2 y ) \;
 t_{ \alpha_2^{'''}} ( \pi_6( q_2' q_2 y ) p_2, E_{ q_2}) \;
   D_{ \alpha_2^{'''}}^{ m_2, m_3} ( \theta_1)\cr
   & - & t_{\alpha_2'} ( \pi_2( q_1' q_2' x ) \pi_5 ( q_2' \tilde q_2 y ), E_{ q_2'}) \;
    G_0( \pi_5 ( q_2' \tilde q_2 y ),q_2')\cr
    & & \sum_{\alpha_2^{'''}} I_{ \alpha_2' \alpha_2^{'''}} ( q_2' \tilde q_2 y ) \;
 t_{ \alpha_2^{'''}} ( \pi_6( q_2' \tilde q_2 y ) \tilde p_2, E_{ \tilde q_2}) \;
   \tilde D_{ \alpha_2^{'''}}^{ m_2, m_3} ( \theta_1)\Bigg] \Bigg].
\label{eq:1.29}
  \end{eqnarray}
   Correspondingly one obtains
\begin{eqnarray}
\lefteqn{\langle p_2' q_2' \alpha_2'| U_{2,a}^{(2)}\equiv }\cr
& &  \sum_{\alpha_1'} \int dx  \int  dq_1' q_1^{'2} \; t_{\alpha_2'} ( p_2' \pi_3 ( q_2'
q_1' x), E_{q_2'}) \;
 G_0( \pi_3 ( q_2' q_1' x),q_2') \cr
 & & H_{ \alpha_2' \alpha_1'}( q_2' q_1' x) \;
\left( 1 + ( -1)^{ l_1' + s_1'} \right) \cr
& &  \int dy \; \Bigg[ t_{ \alpha_1'} ( \pi_4( q_2' q_1'x ) \pi_1 ( q_1' q_2 y ),
E_{q_1'}) \;  G_0(  \pi_1 ( q_1' q_2 y ), q_1') \cr
& & \sum_{\alpha_2'} G_{ \alpha_1' \alpha_2'} ( q_1' q_2 y ) \;
 t_{ \alpha_2'} ( \pi_2 ( q_1' q_2 y ) p_2, E_{ q_2}) \;
  D_{ \alpha_2'}^{ m_2, m_3} ( \theta_1) \cr
& - &  t_{ \alpha_1'} (  \pi_4( q_2' q_1'x ) \pi_1 ( q_1' \tilde q_2 y ), E_{q_1'}) \;
 G_0(  \pi_1 ( q_1' \tilde q_2 y ), q_1') \cr
& & \sum_{\alpha_2'} G_{ \alpha_1' \alpha_2'} ( q_1' \tilde q_2 y ) \;  
 t_{ \alpha_2'} ( \pi_2 ( q_1' \tilde q_2 y ) \tilde p_2, E_{\tilde q_2}) \;
 \tilde D_{ \alpha_2'}^{ m_2, m_3} ( \theta_1) \Bigg] \cr
& - &  \int dx \int d q_2^{'''} (q_2^{'''})^2 \;  t_{\alpha_2'} ( p_2' \pi_5 ( q_2'
q_2^{'''} x ), E_{ q_2'})  \; G_0( \pi_5 ( q_2' q_2^{ '''} x ),q_2') \cr
& & \sum_{\alpha_2^{'''}}  I_{ \alpha_2' \alpha_2^{'''}} ( q_2' q_2^{ '''} x )
\Bigg[ \int dy \;  t_{\alpha_2^{'''}} ( \pi_6( q_2' q_2^{'''} x ) \pi_3 ( q_2^{'''} q_1
y ), E_{ q_2'}) \; G_0( \pi_3 ( q_2^{'''} q_1 y ),q_2^{'''}) \cr
& & \sum_{\alpha_1'}  H_{ \alpha_2^{'''} \alpha_1'}( q_2^{'''} q_1 y ) \; 
t_{\alpha_1'}( \pi_4 ( q_2^{'''} q_1 y ) p_1, E_{q_1}) \;  
C_{\alpha_1'}^{ m_2 + m_3} (\theta_1)\cr
& - &  \int dy \; \Bigg[ t_{\alpha_2^{'''}} ( \pi_6( q_2' q_2^{'''} x ) \pi_5 (
q_2^{'''} q_2 y ), E_{ q_2^{'''}}) \;
G_0( \pi_5 ( q_2^{'''} q_2 y ),q_2^{'''}) \cr
& & \sum_{\alpha_2^{''''}} I_{ \alpha_2^{'''} \alpha_2^{''''}} ( q_2^{'''} q_2 y ) \;
 t_{ \alpha_2^{''''}} ( \pi_6( q_2^{'''} q_2 y) p_2, E_{ q_2}) \;
D_{ \alpha_2^{''''}}^{ m_2, m_3} ( \theta_1) \cr
& - & t_{\alpha_2^{'''}} ( \pi_6( q_2' q_2^{'''} x ) \pi_5 ( q_2^{'''} \tilde q_2 y ),
E_{ q_2^{'''}}) \;
G_0( \pi_5 ( q_2^{'''} \tilde q_2 y ),q_2^{'''})\cr
& & \sum_{\alpha_2^{''''}} I_{ \alpha_2^{'''} \alpha_2^{''''}} ( q_2^{'''} \tilde q_2 y
) \;
 t_{ \alpha_2^{''''}} ( \pi_6( q_2^{'''} \tilde q_2 y ) \tilde p_2, E_{ \tilde q_2}) \;
\tilde D_{ \alpha_2^{'''}}^{ m_2, m_3} ( \theta_1) \Bigg]  \Bigg] \; .
\label{eq:1.30}
\end{eqnarray}

All three-fold integrals in Eqs.~(\ref{eq:1.29}) and (\ref{eq:1.30}) are of the same type: 
two angular integrations, where each one leads to logarithmic singularities in 
the corresponding momenta, one of which is external and the other the intermediate 
integration variable. It is not difficult to see that the intermediate
 momentum  integration over products of logarithms leads to smooth functions in the 
external  momenta. Therefore the third order amplitudes in $t$ can serve as driving terms 
for the application  of the kernels, and thus leading to all higher order amplitudes:
\begin{eqnarray}
\lefteqn{ \langle p_1' q_1' \alpha_1'| U_{1,a}^{(n)}   = }\cr
& &    \left( 1 + ( -1)^{ l_1' + s_1'} \right) \int dx \int dq_2' q_2^{'2} \;
   t_{ \alpha_1} ( p_1' \pi_1 ( q_1' q_2' x), E_{q_1'}) \;
    G_0(  \pi_1 ( q_1' q_2' x), q_1') \cr
& &\sum_{\alpha_2'} G_{ \alpha_1' \alpha_2'} (q_1' q_2' x)  \;
 \langle \pi_2 ( q_1' q_2' x) q_2' \alpha_2'| U_{2,a}^{(n-1)}\cr
\lefteqn{ \langle p_2' q_2' \alpha_2'| U_{2,a}^{(n)} = } \cr
& &  \int dx  \int  dq_1' q_1^{'2} \; t_{\alpha_2'} ( p_2' \pi_3 ( q_2' q_1' x),
E_{q_2'}) \; G_0( \pi_3 ( q_2' q_1' x),q_2') \cr
  & &  \sum_{\alpha_1'} H_{ \alpha_2' \alpha_1'}( q_2' q_1' x) \;
 \langle \pi_4 ( q_2' q_1' x) q_1' \alpha_1' | U_{1,a}^{(n-1)}\cr
  & - &  \int dx \int dq_2^{'''} q_2^{'''2} \; t_{\alpha_2'} ( p_2' \pi_5 ( q_2'
q_2^{'''} x), E_{ q_2'}) \;
    G_0( \pi_5 ( q_2' q_2^{'''} x),q_2') \cr
  & & \sum_{\alpha_2^{'''}}  I_{ \alpha_2' \alpha_2^{'''}} ( q_2' q_2^{'''} x) \; \langle
\pi_6( q_2' q_2^{'''} x) q_2^{'''} \alpha_2^{'''} | U_{2,a}^{(n-1)} \; ,
\label{eq:1.31}
  \end{eqnarray}
with $ n= 3,4,\cdots$.

The resulting series $ \sum_{n=3}^{\infty}\langle p_1' q_1' \alpha_1'| U_{1,a}^{(n)}$
and  $ \sum_{n=3}^{\infty}\langle p_2' q_2' \alpha_2'| U_{2,a}^{(n)}$ 
can safely be summed e.g. via Pad\'e summation. For the corresponding three-nucleon 
amplitudes the above considerations were made in Ref.~\cite{vanMeijgaard:1992zz}.

\section{Finite Rank Forces}
\label{section5}

So far, Faddeev type studies of light nuclei treating the discrete structures
have been based on
finite rank forces~\cite{Eskandarian:1992zz}.
Therefore, it appears useful to also formulate the $nn\alpha$ system in the continuum 
in this fashion. 
For the sake of a simple notation we choose a rank-1 separable t-matrix,
\begin{eqnarray}
t_{\alpha} ( p p', E_q) = h_{\alpha} (p) \tau_{\alpha} ( q)h_{\alpha} (p') \; .
\label{eq:5.1}
 \end{eqnarray}
Then Eqs.~(\ref{eq:1.18}) take the form
 \begin{eqnarray}
\lefteqn{ \langle p_1' q_1' \alpha_1'| U_{1,a}  = } \cr
& & \frac{\delta( q_1' - q_1)}{ q_1^2}\;  h_{\alpha_1'} (p_1') \tau_{\alpha_1'}
(E_{q_1}) h_{\alpha_1'} (p_1) \;
 C_{\alpha_1'}^{ m_2 + m_3} ( \theta_1) \cr
 & + &  \left( 1 + ( -1)^{ l_1' + s_1'}\right) 
\int dx \int dq_2' q_2^{'2} \sum_{\alpha_2'} h_{\alpha_1'} (p_1') \tau_{\alpha_1'}
(E_{q_1'}) 
h_{\alpha_1'} (\pi_1 ( q_1' q_2' x)) \cr 
& &  G_0(  \pi_1 ( q_1' q_2' x), q_1') \; 
   G_{ \alpha_1' \alpha_2'} ( q_1' q_2' x) \;  \langle \pi_2 ( q_1' q_2' x) q_2'
\alpha_2'| U_{2,a}\cr
&\equiv&  h_{\alpha_1'} (p_1') Z_{\alpha_1'} ( q_1'),
\label{eq:5.2}
 \end{eqnarray}
where the new unknown single variable amplitude is
\begin{eqnarray}
\lefteqn{ Z_{\alpha_1'} ( q_1') = } \cr
& & \frac{\delta( q_1' - q_1)}{ q_1^2} \;  \tau_{\alpha_1'} (E_{q_1}) h_{\alpha_1'}
(p_1) \;
 C_{\alpha_1'}^{ m_2 + m_3} ( \theta_1)\cr
 & + &  \left( 1 + ( -1)^{ l_1' + s_1'}\right) \int dx \int dq_2' q_2^{'2} 
\sum_{\alpha_2'} \tau_{\alpha_1'} (E_{q_1'}) h_{\alpha_1'} (\pi_1 ( q_1' q_2' x)) \cr 
& & G_0(  \pi_1 ( q_1' q_2' x), q_1') \;
   G_{ \alpha_1' \alpha_2'} ( q_1' q_2' x) \; \langle \pi_2 ( q_1' q_2' x) q_2'
\alpha_2'| U_{2,a}.
\label{eq:5.2a}
\end{eqnarray}
Similarly, the second equation, Eq.~(\ref{eq:1.18}) becomes
\begin{eqnarray}
\lefteqn{ \langle p_2' q_2' \alpha_2'| U_{2,a} = } \cr
& &  \frac{ \delta( q_2' - q_2)}{ q_2^2} \; h_{\alpha_2'} (p_2') \tau_{\alpha_2'}
(E_{q_2}) h_{\alpha_2'} (p_2) \;
  D_{ \alpha_2'}^{ m_2, m_3} ( \theta_1)\cr
& + & \sum_{\alpha_1'} \int dx  \int  dq_1' q_1^{'2} \;  h_{\alpha_2'} (p_2')
\tau_{\alpha_2'} (E_{q_2'}) 
h_{\alpha_2'} (\pi_3 ( q_2' q_1' x)) \cr
& &  G_0( \pi_3 ( q_2' q_1' x),q_2') \;
 H_{ \alpha_2' \alpha_1'}( q_2' q_1' x) \; \langle \pi_4 ( q_2' q_1' x) q_1' \alpha_1' |
U_{1,a}\cr
  & - &  \sum_{\alpha_2^{'''}}  \int dx \int dq_2^{'''} q_2^{'''2} \;  h_{\alpha_2'}
(p_2') \tau_{\alpha_2'}
(E_{q_2'}) h_{\alpha_2'} (\pi_5 ( q_2' q_2^{'''} x)) \cr
& & G_0( \pi_5 ( q_2' q_2^{'''} x),q_2')\;
  I_{ \alpha_2' \alpha_2^{'''}} ( q_2' q_2^{'''} x) \; 
\langle \pi_6( q_2' q_2^{'''} x) q_2^{'''} \alpha_2^{'''} | U_{2,a}\cr
  &\equiv& h_{\alpha_2'} (p_2')V_{\alpha_2'} ( q_2') \; ,
\label{eq:5.3}
 \end{eqnarray}
  with
\begin{eqnarray}
\lefteqn{ V_{\alpha_2'} ( q_2') = } \cr
& & \frac{ \delta( q_2' - q_2)}{ q_2^2}  \tau_{\alpha_2'} (E_{q_2}) h_{\alpha_2'} (p_2)
\; 
 D_{ \alpha_2'}^{ m_2, m_3} ( \theta_1) \cr
& + & \sum_{\alpha_1'} \int dx  \int  dq_1' q_1^{'2} \; \tau_{\alpha_2'} (E_{q_2'})
h_{\alpha_2'} (\pi_3 ( q_2' q_1'
x)) \cr
& & G_0( \pi_3 ( q_2' q_1' x),q_2')\;
 H_{ \alpha_2' \alpha_1'}( q_2' q_1' x) \;  \langle \pi_4 ( q_2' q_1' x) q_1' \alpha_1'
| U_{1,a}\cr
  & - &  \sum_{\alpha_2^{'''}}  \int dx \int dq_2^{'''} q_2^{'''2} \; \tau_{\alpha_2'}
(E_{q_2'}) 
h_{\alpha_2'} (\pi_5 ( q_2' q_2^{'''} x)) \cr 
& & G_0( \pi_5 ( q_2' q_2^{'''} x),q_2')\;
  I_{ \alpha_2' \alpha_2^{'''}} ( q_2' q_2^{'''} x)\;  
\langle \pi_6( q_2' q_2^{'''} x) q_2^{'''} \alpha_2^{'''} | U_{2,a} \; .
\label{eq:5.3a}
\end{eqnarray}
Then we insert the functions $ Z$ and $V$ under the integrals
\begin{eqnarray}
\lefteqn{ Z_{\alpha_1'} ( q_1') = } \cr
& & \frac{\delta( q_1' - q_1)}{ q_1^2}  \tau_{\alpha_1'} ( q_1) h_{\alpha_1'} (p_1)
 C_{\alpha_1'}^{ m_2 + m_3} ( \theta_1)\cr
 & + &  ( 1 + ( -1)^{ l_1' + s_1'}) \int dx \int dq_2' q_2^{'2} \sum_{\alpha_2'}
\tau_{\alpha_1'} (E_{q_1'}) h_{\alpha_1'} 
(\pi_1 ( q_1' q_2' x)) \cr
& & G_0(  \pi_1 ( q_1' q_2' x), q_1')\;
   G_{ \alpha_1' \alpha_2'} ( q_1' q_2' x) h_{\alpha_2'} (\pi_2 ( q_1' q_2' x) )
V_{\alpha_2'} ( q_2'),
\label{eq:5.4a}
\end{eqnarray}
and
\begin{eqnarray}
\lefteqn{  V_{\alpha_2'} ( q_2')  = } \cr
& & \frac{ \delta( q_2' - q_2)}{ q_2^2}  \tau_{\alpha_2'} ( q_2) h_{\alpha_2'} (p_2) \; 
 D_{ \alpha_2'}^{ m_2, m_3} ( \theta_1)\cr
& + & \sum_{\alpha_1'} \int dx  \int  dq_1' q_1^{'2} \; \tau_{\alpha_2'} (E_{q_2'})
h_{\alpha_2'} (\pi_3 ( q_2' q_1'
x)) \cr
& &  G_0( \pi_3 ( q_2' q_1' x),q_2')\;
   H_{ \alpha_2' \alpha_1'}( q_2' q_1' x)h_{\alpha_1'} (\pi_4 ( q_2' q_1' x))
Z_{\alpha_1'} ( q_1') \cr
  & - &  \sum_{\alpha_2^{'''}}  \int dx \int dq_2^{'''} q_2^{'''2} \;  \tau_{\alpha_2'}
(E_{q_2'}) 
h_{\alpha_2'} (\pi_5 ( q_2' q_2^{'''} x)) \cr
& &     G_0( \pi_5 ( q_2' q_2^{'''} x),q_2')\;
    I_{ \alpha_2' \alpha_2^{'''}} ( q_2' q_2^{'''} x) h_{\alpha_2^{'''}} (\pi_6( q_2'
q_2^{'''} x)) 
V_{\alpha_2^{'''}} ( q_2^{'''}).
\label{eq:5.4b}
\end{eqnarray}
These two equations, Eqs.~(\ref{eq:5.4a}) and (\ref{eq:5.4b}), form as set of  coupled
one-dimensional integral equations.
The low order iterations exhibits the same features as discussed in detail in the
previous section. Thus, we write
\begin{eqnarray}
Z_{\alpha_1'} ( q_1') = Z_{\alpha_1'}^{(0)} ( q_1') + Z_{\alpha_1'}^{(1)}( q_1') + Z_{\alpha_1'}^{(2)} ( q_1') + Z_{\alpha_1'}^{(3)} ( q_1') + \cdots
\label{eq:5.5a}
\end{eqnarray}
and
 \begin{eqnarray}
V_{\alpha_2'} ( q_2') = V_{\alpha_2'}^{(0)} ( q_2') + V_{\alpha_2'}^{(1)}( q_2') +
V_{\alpha_2'}^{(2)} ( q_2') + V_{\alpha_2'}^{(3)} ( q_2') + \cdots  \; .
\label{eq:5.5b}
\end{eqnarray}
From Eqs.~(\ref{eq:5.4a}) and (\ref{eq:5.4b}) we can read off the different orders.
For the lowest order we obtain,
\begin{eqnarray}
 Z_{\alpha_1'}^{(0)} ( q_1') = \frac{\delta( q_1' - q_1)}{ q_1^2} \;  \tau_{\alpha_1'}(
q_1)\; h_{\alpha_1'} (p_1)\; C_{\alpha_1'}^{ m_2 + m_3} ( \theta_1)
\label{eq:5.6a}
 \end{eqnarray}
 and
 \begin{eqnarray}
 V_{\alpha_2'}^{(0)}(q_2')& =& 
 \frac{ \delta( q_2' - q_2)}{ q_2^2} \; \tau_{\alpha_2'} ( q_2)\; h_{\alpha_2'} (p_2)
\; D_{ \alpha_2'}^{ m_2, m_3} ( \theta_1) \cr
& -&  \frac{ \delta( q_2' - \tilde q_2)}{ {\tilde q_2}^2} \; \tau_{\alpha_2'} ( \tilde
    q_2) \; h_{\alpha_2'} (\tilde p_2 \;)
  \tilde D_{ \alpha_2'}^{ m_2, m_3} ( \theta_1)  \; .
\label{eq:5.6b}
\end{eqnarray}
The second order is given by
\begin{eqnarray}
\lefteqn{Z_{\alpha_1'}^{(1)} ( q_1') = }\cr
& &  \left( 1 + ( -1)^{ l_1' + s_1'}\right) \int dx \int dq_2' q_2^{'2} \sum_{\alpha_2'} 
\tau_{\alpha_1'} (E_{q_1'})\; h_{\alpha_1'} (\pi_1 ( q_1' q_2' x))\; G_0(  \pi_1 ( q_1'
q_2' x), q_1')\cr
& & G_{ \alpha_1' \alpha_2'} ( q_1' q_2' x) \; h_{\alpha_2'} (\pi_2 ( q_1' q_2' x) )\;  
V_{\alpha_2'}^{(0) }( q_2')\cr
 & = & \left( 1 + ( -1)^{ l_1' + s_1'}\right) \int dx \int dq_2' q_2^{'2}
\sum_{\alpha_2'} 
\tau_{\alpha_1'} (E_{q_1'}) \;  h_{\alpha_1'} (\pi_1 ( q_1' q_2' x)) \;
   G_0(  \pi_1 ( q_1' q_2' x), q_1')\cr
 & & \sum_{\alpha_2'} G_{ \alpha_1' \alpha_2'} ( q_1' q_2' x) \; h_{\alpha_2'} ( \pi_2 (
q_1' q_2' x) ) \Bigg[ \frac{ \delta( q_2' - q_2)}{ q_2^2} \;  \tau_{\alpha_2'} ( q_2) \;
   h_{\alpha_2'} (p_2)  \; D_{ \alpha_2'}^{ m_2, m_3} ( \theta_1)\cr
& - &  \frac{ \delta( q_2' - \tilde q_2)}{ {\tilde q_2}^2} \; \tau_{\alpha_2'} ( \tilde
q_2) \; h_{\alpha_2'} (\tilde p_2) \;
   \tilde D_{ \alpha_2'}^{ m_2, m_3} ( \theta_1) \Bigg]\cr
 & = & \left( 1 + ( -1)^{ l_1' + s_1'}\right) \; \tau_{\alpha_1'} ( q_1') \int dx \;
   \Bigg[h_{\alpha_1'} (\pi_1 ( q_1' q_2 x)) \; G_0(  \pi_1 ( q_1' q_2 x), q_1')\cr
& &\sum_{\alpha_2'} G_{ \alpha_1' \alpha_2'} ( q_1' q_2 x) \; h_{\alpha_2'} (\pi_2 (
q_1' q_2 x) ) \; \tau_{\alpha_2'} ( q_2)\; h_{\alpha_2'} (p_2) \;
   D_{ \alpha_2'}^{ m_2, m_3} ( \theta_1)\cr
& - & h_{\alpha_1'} (\pi_1 ( q_1' \tilde q_2 x)) \; G_0(\pi_1 ( q_1' \tilde q_2 x), q_1')\cr
  & & \sum_{\alpha_2'} G_{ \alpha_1' \alpha_2'} ( q_1' \tilde q_2 x) \; 
h_{\alpha_2'} (\pi_2 ( q_1' \tilde q_2 x) ) \;  \tau_{\alpha_2'} ( \tilde q_2) \;  
h_{\alpha_2'} (\tilde p_2) \; \tilde D_{ \alpha_2'}^{ m_2, m_3} ( \theta_1)\Bigg]\cr
  & = & \left( 1 + ( -1)^{ l_1' + s_1'}\right) \; \tau_{\alpha_1'} ( q_1') \sum_{\alpha_2'}
 \Bigg[\tau_{\alpha_2'} ( q_2)\;  h_{\alpha_2'} (p_2)\;  D_{ \alpha_2'}^{ m_2, m_3} ( \theta_1)\cr
& & \int dx \; h_{\alpha_1'}(\pi_1 ( q_1' q_2 x))\;  G_0(  \pi_1 ( q_1' q_2 x), q_1')\;
G_{ \alpha_1' \alpha_2'} ( q_1' q_2 x)\;  h_{\alpha_2'} (\pi_2 ( q_1' q_2 x))\cr
  & - &   \tau_{\alpha_2'} ( \tilde q_2)\;  h_{\alpha_2'} (\tilde p_2)\;  \tilde D_{ \alpha_2'}^{ m_2, m_3} ( \theta_1)\cr
  & & \int dx \; h_{\alpha_1'} (\pi_1 ( q_1' \tilde q_2 x))\;  G_0(  \pi_1 ( q_1' \tilde
q_2 x), q_1') \; G_{ \alpha_1' \alpha_2'} ( q_1' \tilde q_2 x)
\;   h_{\alpha_2'} (\pi_2 ( q_1' \tilde q_2 x))\Bigg]
\label{eq:5.7}
  \end{eqnarray}
 and
\begin{eqnarray}
\lefteqn{ V_{\alpha_2'}^{(1)} ( q_2') = } \cr
& &  \int dx  \int  dq_1' q_1^{'2} \; \tau_{\alpha_2'} ( q_2') \; h_{\alpha_2'} (\pi_3 (
q_2' q_1' x)) \;
  G_0( \pi_3 ( q_2' q_1' x),q_2')\cr
& & \sum_{\alpha_1'} H_{ \alpha_2' \alpha_1'}( q_2' q_1' x) \; h_{\alpha_1'} (\pi_4 (
q_2' q_1' x)) \; Z_{\alpha_1'}^{(0)} ( q_1')\cr
 &-&   \int dx \int dq_2^{'''} q_2^{'''2} \;  \tau_{\alpha_2'} ( q_2') \; h_{\alpha_2'}
(\pi_5 ( q_2' q_2^{'''} x)) \; G_0( \pi_5 ( q_2' q_2^{'''} x),q_2')\cr
& & \sum_{\alpha_2^{'''}} I_{ \alpha_2' \alpha_2^{'''}} ( q_2' q_2^{'''} x) \; 
h_{\alpha_2^{'''}} (\pi_6( q_2' q_2^{'''} x)) \; V_{\alpha_2^{'''}}^{(0)} ( q_2^{'''})\cr
  & = &  \int dx  \int  dq_1' q_1^{'2} \; \tau_{\alpha_2'} ( q_2')\;  h_{\alpha_2'} (\pi_3 ( q_2' q_1' x)) G_0( \pi_3 ( q_2' q_1' x),q_2')\cr
  & &\sum_{\alpha_1'} H_{ \alpha_2' \alpha_1'}( q_2' q_1' x) \; h_{\alpha_1'} (\pi_4 (
q_2' q_1' x)) \; \frac{\delta( q_1' - q_1)}{ q_1^2} \;  \tau_{\alpha_1'} ( q_1)
\;    h_{\alpha_1'} (p_1)\;  C_{\alpha_1'}^{ m_2 + m_3} ( \theta_1)\cr
 &-&  \int dx \int dq_2^{'''} q_2^{'''2} \;  \tau_{\alpha_2'} ( q_2')\;  h_{\alpha_2'}
(\pi_5 ( q_2' q_2^{'''} x)) \; G_0( \pi_5 ( q_2' q_2^{'''} x),q_2')\cr
  & & \sum_{\alpha_2^{'''}} I_{ \alpha_2' \alpha_2^{'''}} ( q_2' q_2^{'''} x) \;
 h_{\alpha_2^{'''}} (\pi_6( q_2' q_2^{'''} x))
  \Bigg[ \frac{ \delta( q_2^{'''} - q_2)}{ q_2^2}\;  \tau_{\alpha_2{'''}} ( q_2)\;
h_{\alpha_2^{'''}} (p_2)\;  D_{ \alpha_2^{'''}}^{ m_2, m_3} ( \theta_1)\cr
  &-& \frac{ \delta( q_2^{'''} - \tilde q_2)}{ {\tilde q_2}^2} \; \tau_{\alpha_2{'''}} (
\tilde q_2)\; h_{\alpha_2^{'''}} (\tilde p_2)
 \; \tilde D_{ \alpha_2^{'''}}^{ m_2, m_3} ( \theta_1)\Bigg]\cr
  & = &  \int dx  \; \tau_{\alpha_2'} ( q_2')\; h_{\alpha_2'} (\pi_3 ( q_2' q_1 x))
\;  G_0( \pi_3 ( q_2' q_1 x),q_2')\cr
  & & \sum_{\alpha_1'} H_{ \alpha_2' \alpha_1'}( q_2' q_1 x)\; h_{\alpha_1'} (\pi_4 (
q_2' q_1 x)) \;  \tau_{\alpha_1'} ( q_1)\; h_{\alpha_1'} (p_1)
\; C_{\alpha_1'}^{ m_2 + m_3} ( \theta_1)\cr
 &-&   \int dx  \Bigg[ \tau_{\alpha_2'} ( q_2') \; h_{\alpha_2'} (\pi_5 ( q_2' q_2 x))
   \;   G_0( \pi_5 ( q_2' q_2 x),q_2')\cr
  & & \sum_{\alpha_2^{'''}} I_{ \alpha_2' \alpha_2^{'''}} ( q_2' q_2 x) \;
h_{\alpha_2^{'''}} (\pi_6( q_2' q_2 x)) \;   \tau_{\alpha_2{'''}} ( q_2) \; 
h_{\alpha_2^{'''}} (p_2) \; D_{ \alpha_2^{'''}}^{ m_2, m_3} ( \theta_1)\cr
    & - & \tau_{\alpha_2'} ( q_2')\; h_{\alpha_2'} (\pi_5 ( q_2' \tilde q_2 x))
   \;  G_0( \pi_5 ( q_2' \tilde q_2 x),q_2')\cr
  & &  I_{ \alpha_2' \alpha_2^{'''}} ( q_2' \tilde q_2 x)\; h_{\alpha_2^{'''}} (\pi_6(
q_2' \tilde q_2 x)) \;  \tau_{\alpha_2{'''}} ( \tilde q_2)\; h_{\alpha_2^{'''}} (p_2)\; 
\tilde D_{ \alpha_2^{'''}}^{ m_2, m_3} ( \theta_1)\Bigg]\cr
  & = & \tau_{\alpha_2'} ( q_2')\sum_{\alpha_1'}  \tau_{\alpha_1'} ( q_1) \;
h_{\alpha_1'} (p_1) \; C_{\alpha_1'}^{ m_2 + m_3} ( \theta_1)\cr
 & & \int dx  \;  h_{\alpha_2'} (\pi_3 ( q_2' q_1 x)) \; G_0( \pi_3 ( q_2' q_1 x),q_2')
\; H_{ \alpha_2' \alpha_1'}( q_2' q_1 x) \; h_{\alpha_1'} (\pi_4 ( q_2'x))\cr
 &-& \tau_{\alpha_2'} ( q_2')\sum_{\alpha_2^{'''}} \Bigg[ \tau_{\alpha_2{'''}} ( q_2)\;
h_{\alpha_2^{'''}} (p_2)\; D_{ \alpha_2^{'''}}^{ m_2, m_3} ( \theta_1)\cr
   & &  \int dx\; h_{\alpha_2'} (\pi_5 ( q_2' q_2 x))\; G_0( \pi_5( q_2' q_2 x),q_2')\; 
I_{ \alpha_2' \alpha_2^{'''}}( q_2' q_2 x) \; h_{\alpha_2^{'''}} (\pi_6( q_2' q_2 x))\cr
   & - & \tau_{\alpha_2{'''}} ( \tilde q_2) \; h_{\alpha_2^{'''}} (\tilde p_2) \; 
\tilde D_{ \alpha_2^{'''}}^{ m_2, m_3} ( \theta_1)\cr
   & &  \int dx \;  h_{\alpha_2'} (\pi_5 ( q_2' \tilde q_2 x)) \;  G_0( \pi_5( q_2'
\tilde q_2 x),q_2') \; I_{ \alpha_2' \alpha_2^{'''}}( q_2' \tilde q_2 x)\cr
   & &  h_{\alpha_2^{'''}} (\pi_6( q_2' \tilde q_2 x))\Bigg] \; .
\label{eq:5.8}
  \end{eqnarray}
As for general forces the x-integration leads to logarithmic singularities 
in the external momenta.
The next order, however, gives smooth functions
\begin{eqnarray}
\lefteqn{ Z_{\alpha_1'}^{(2)} ( q_1') = } \cr 
& & \left( 1 + ( -1)^{ l_1' + s_1'}\right) \; \tau_{\alpha_1'} ( q_1') \int dx \int
dq_2' q_2^{'2} \;
  h_{\alpha_1'} (\pi_1 ( q_1' q_2' x)) \;  G_0(  \pi_1 ( q_1' q_2' x), q_1')\cr
  & & \sum_{\alpha_2'}G_{ \alpha_1' \alpha_2'} ( q_1' q_2' x) \;  h_{\alpha_2'} (\pi_2 (
q_1' q_2' x) ) \; V_{\alpha_2'}^{(1)} ( q_2')\cr
  & = & \left( 1 + ( -1)^{ l_1' + s_1'}\right) \; \tau_{\alpha_1'} ( q_1') \int dq_2' q_2^{'2}\cr
  & &  \int dx \; h_{\alpha_1'} (\pi_1 ( q_1' q_2' x)) \; G_0(  \pi_1 ( q_1' q_2' x), q_1') \sum_{\alpha_2'} G_{ \alpha_1' \alpha_2'} ( q_1' q_2' x)
 \;   h_{\alpha_2'} (\pi_2 ( q_1' q_2' x) ) \;\tau_{\alpha_2'} ( q_2')\cr
& & \Bigg[ \sum_{\alpha_1^{''}}  \tau_{\alpha_1^{''}} ( q_1) \; h_{\alpha_1^{''}} (p_1) \; C_{\alpha_1^{''}}^{ m_2 + m_3} ( \theta_1)\cr
 & & \int dy \;  h_{\alpha_2'} (\pi_3 ( q_2' q_1 y ))\; G_0( \pi_3 ( q_2' q_1 y
),q_2')\; H_{ \alpha_2' \alpha_1'}( q_2' q_1 y ) \; h_{\alpha_1'} (\pi_4 ( q_2'y ))\cr
 &-& \sum_{\alpha_2^{'''}} \Bigg[ \tau_{\alpha_2{'''}} ( q_2) \; h_{\alpha_2^{'''}}
(p_2) \; D_{ \alpha_2^{'''}}^{ m_2, m_3} ( \theta_1)\cr
   & &  \int dy \; h_{\alpha_2'} (\pi_5 ( q_2' q_2 y )) \; G_0( \pi_5( q_2' q_2 y
),q_2') \; I_{ \alpha_2' \alpha_2^{'''}}( q_2' q_2 y )
\;   h_{\alpha_2^{'''}} (\pi_6( q_2' q_2 y ))\cr
   & - & \tau_{\alpha_2{'''}} ( \tilde q_2)\;  h_{\alpha_2^{'''}} (\tilde p_2)\;  \tilde D_{ \alpha_2^{'''}}^{ m_2, m_3} ( \theta_1)\cr
   & &  \int dy\;  h_{\alpha_2'} (\pi_5 ( q_2' \tilde q_2 y ))\;  G_0( \pi_5( q_2' \tilde q_2 y ),q_2')
\;    I_{ \alpha_2' \alpha_2^{'''}}( q_2' \tilde q_2 y )\;  h_{\alpha_2^{'''}} (\pi_6(
q_2' \tilde q_2 y ))\Bigg]\Bigg] 
\label{eq:5.9}
\end{eqnarray}
 and
 \begin{eqnarray}
\lefteqn{ V_{\alpha_2'}^{(2)} ( q_2') = } \cr
& &  \int dx  \int  dq_1' q_1^{'2} \; \tau_{\alpha_2'} ( q_2') \; h_{\alpha_2'} (\pi_3 ( q_2' q_1' x))
 \;  G_0( \pi_3 ( q_2' q_1' x),q_2')\cr
  & &\sum_{ \alpha_1'} H_{ \alpha_2' \alpha_1'}( q_2' q_1' x) \; h_{\alpha_1'} (\pi_4 (
q_2' q_1' x)) \; Z_{\alpha_1'}^{(1)} ( q_1')\cr
 &-&   \int dx \int dq_2^{'''} q_2^{'''2}  \; \tau_{\alpha_2'} ( q_2') \; h_{\alpha_2'} (\pi_5 ( q_2' q_2^{'''} x))
 \;     G_0( \pi_5 ( q_2' q_2^{'''} x),q_2')\cr
  & & \sum_{\alpha_2^{'''}} I_{ \alpha_2' \alpha_2^{'''}} ( q_2' q_2^{'''} x) \;
h_{\alpha_2^{'''}} (\pi_6( q_2' q_2^{'''} x)) \; V_{\alpha_2^{'''}}^{(1)} ( q_2^{'''})\cr
   & = & \tau_{\alpha_2'} ( q_2') \int  dq_1' q_1^{'2} \cr
  & &  \int dx   \;  h_{\alpha_2'} (\pi_3 ( q_2' q_1' x))
 \;  G_0( \pi_3 ( q_2' q_1' x),q_2') \sum_{\alpha_1'} H_{ \alpha_2' \alpha_1'}( q_2'
q_1' x)  \; h_{\alpha_1'} (\pi_4 ( q_2' q_1' x))\cr
    & & \left( 1 + ( -1)^{ l_1' + s_1'}\right) \; \tau_{\alpha_1'} ( q_1') \sum_{\alpha_2^{''}}
\; \Bigg[  \tau_{\alpha_2^{''}} ( q_2) \;  h_{\alpha_2^{''}} (p_2) \;  D_{ \alpha_2^{''}}^{ m_2, m_3} ( \theta_1)\cr
  & & \int dy \;  h_{\alpha_1'} (\pi_1 ( q_1' q_2 y)) \;  G_0(  \pi_1 ( q_1' q_2 y ),
q_1') \;  G_{ \alpha_1' \alpha_2^{''}} ( q_1' q_2 y )
 \;   h_{\alpha_2^{''}} (\pi_2 ( q_1' q_2 y ))\cr
  & - &   \tau_{\alpha_2^{''}} ( \tilde q_2) \;  h_{\alpha_2^{''}} (\tilde p_2) \;  \tilde D_{ \alpha_2^{''}}^{ m_2, m_3} ( \theta_1)\cr
  & & \int dy \;   h_{\alpha_1'} (\pi_1 ( q_1' \tilde q_2 y )) \;  G_0(  \pi_1 ( q_1'
\tilde q_2 y ), q_1') \;  G_{ \alpha_1' \alpha_2^{''}} ( q_1' \tilde q_2 y )
 \;   h_{\alpha_2^{''}} (\pi_2 ( q_1' \tilde q_2 y ))]\cr
    & - & \tau_{\alpha_2'} ( q_2') \int dq_2^{'''} q_2^{'''2} \cr
  & &   \int dx  \;  h_{\alpha_2'} (\pi_5 ( q_2' q_2^{'''} x)) \;  G_0( \pi_5 ( q_2' q_2^{'''} x),q_2')
   \sum_{\alpha_2^{'''}} I_{ \alpha_2' \alpha_2^{'''}} ( q_2' q_2^{'''} x) \;  h_{\alpha_2^{'''}} (\pi_6( q_2' q_2^{'''} x))\cr
 & &   \Bigg[ \tau_{\alpha_2^{'''}} ( q_2^{'''}) \sum_{\alpha_1'}  \tau_{\alpha_1'} (
q_1) \; h_{\alpha_1'} (p_1) C_{\alpha_1'}^{ m_2 + m_3} ( \theta_1)\cr
 & & \int dy  \;  h_{\alpha_2^{'''}} (\pi_3 ( q_2^{'''} q_1 y )) \; G_0( \pi_3 (
q_2^{'''} q_1 y ),q_2^{'''}) \; H_{ \alpha_2^{'''} \alpha_{1}'}( q_2^{'''} q_1 y ) \; h_{\alpha_1'} (\pi_4 ( q_2^{'''}y ))\cr
 &-& \tau_{\alpha_2^{'''}} ( q_2^{'''})\sum_{\alpha_2^{''''}} \Bigg[
\tau_{\alpha_2{''''}} ( q_2)\;  h_{\alpha_2^{''''}} (p_2)
\;  D_{ \alpha_2^{''''}}^{ m_2, m_3} ( \theta_1)\cr
   & &  \int dy\;  h_{\alpha_2^{'''}} (\pi_5 ( q_2^{'''} q_2 y ))\;  G_0( \pi_5(
q_2^{'''} q_2 y ),q_2^{'''})\; I_{ \alpha_2^{'''} \alpha_2^{''''}}( q_2^{'''} q_2 y )
\;    h_{\alpha_2^{''''}} (\pi_6( q_2^{'''} q_2 y ))\cr
   & - & \tau_{\alpha_2{''''}} ( \tilde q_2)\;  h_{\alpha_2^{''''}} (p_2)\;  \tilde D_{ \alpha_2^{''''}}^{ m_2, m_3} ( \theta_1)\cr
   & &  \int dy\;  h_{\alpha_2^{'''}} (\pi_5 ( q_2^{'''} \tilde q_2 y ))\;  G_0( \pi_5( q_2^{'''} \tilde q_2 y ),q_2^{'''})
\;    I_{ \alpha_2^{'''} \alpha_2^{''''}}( q_2^{'''} \tilde q_2 y )\cr
   & &  h_{\alpha_2^{''''}} (\pi_6( q_2^{'''} \tilde q_2 y ))\Bigg] \Bigg] \; .
\label{eq:5.10}
 \end{eqnarray}
As shown in the previous section, there appear three-fold integrals. Two of them are
over angles, leading to logarithmic singularities, and which are then eliminated by the
third intermediate momentum integral.

\noindent
Thus we end up starting with $ n=3 $:
\begin{eqnarray}
\lefteqn{ Z_{\alpha_1'}^{(n)} ( q_1') = }\cr
& &  \left( 1 + ( -1)^{ l_1' + s_1'}\right) \int dx \int dq_2' q_2^{'2} \;
 \tau_{\alpha_1'} ( q_1') \; h_{\alpha_1'} (\pi_1 ( q_1' q_2' x)) \; 
G_0(  \pi_1 ( q_1' q_2' x), q_1')\cr
& &\sum_{\alpha_2'} G_{ \alpha_1' \alpha_2'} ( q_1' q_2' x) \; h_{\alpha_2'} (\pi_2 (
q_1' q_2' x) ) \; V_{\alpha_2'}^{(n-1) }( q_2')\cr
& = & \left( 1 + ( -1)^{ l_1' + s_1'}\right) \; \tau_{\alpha_1'} ( q_1')\int dq_2'
q_2^{'2} \; \int dx \; h_{\alpha_1'} (\pi_1 ( q_1' q_2' x)) \; G_0(  \pi_1 ( q_1' q_2' x), q_1')\cr
&& \sum_{\alpha_2'}G_{ \alpha_1' \alpha_2'} ( q_1' q_2' x) \; h_{\alpha_2'} (\pi_2 (
q_1' q_2' x) ) \; V_{\alpha_2'}^{(n-1) }( q_2') \; .
\end{eqnarray}
\label{eq:5.11}
Correspondingly,
\begin{eqnarray}
\lefteqn{ V_{\alpha_2'}^{(n) }( q_2')  =  } \cr
& &   \int dx  \int  dq_1' q_1^{'2} \; \tau_{\alpha_2'} ( q_2') \; 
h_{\alpha_2'} (\pi_3 ( q_2' q_1' x)) \; G_0( \pi_3 ( q_2' q_1' x),q_2')\cr
  & &\sum_{\alpha_1'} H_{ \alpha_2' \alpha_1'}( q_2' q_1' x)\; h_{\alpha_1'} (\pi_4 (
q_2' q_1' x))\; Z_{\alpha_1'}^{(n-1)} ( q_1')\cr
 &-&   \int dx \int dq_2^{'''} q_2^{'''2} \; \tau_{\alpha_2'} ( q_2')\; h_{\alpha_2'} (\pi_5 ( q_2' q_2^{'''} x))
\;     G_0( \pi_5 ( q_2' q_2^{'''} x),q_2')\cr
  & & \sum_{\alpha_2^{'''}} I_{ \alpha_2' \alpha_2^{'''}} ( q_2' q_2^{'''} x)\;
h_{\alpha_2^{'''}} (\pi_6( q_2' q_2^{'''} x))\; V_{\alpha_2^{'''}}^{(n-1)} ( q_2^{'''})\cr
  & = & \tau_{\alpha_2'} ( q_2') \int  dq_1' q_1^{'2}
 \int dx\; h_{\alpha_2'} (\pi_3 ( q_2' q_1' x))\; G_0( \pi_3 ( q_2' q_1' x),q_2')\cr
  & & \sum_{\alpha_1'}H_{ \alpha_2' \alpha_1'}( q_2' q_1' x)\; h_{\alpha_1'} (\pi_4 (
q_2' q_1' x))\; Z_{\alpha_1'}^{(n-1)} ( q_1')\cr
  & - & \tau_{\alpha_2'} ( q_2')\int dq_2^{'''} q_2^{'''2}  \int dx \; h_{\alpha_2'}
(\pi_5 ( q_2' q_2^{'''} x)) \; G_0( \pi_5 ( q_2' q_2^{'''} x),q_2')\cr
  & & \sum_{\alpha_2^{'''}} I_{ \alpha_2' \alpha_2^{'''}} ( q_2' q_2^{'''} x) \;
h_{\alpha_2^{'''}} (\pi_6( q_2' q_2^{'''} x)) \; V_{\alpha_2^{'''}}^{(n-1)} ( q_2^{'''})
\; .
\label{eq:5.12}
  \end{eqnarray}
Again, the singular integrals can be rewritten according to the method given in
Appendix~\ref{appendixb}.

\section{The Unitarity Relations}
\label{section6}

The scattering states $\Psi_0^{(+)}$ depend on the initial state quantum numbers
\begin{eqnarray}
\Psi_0^{(+)} \equiv \Psi_{ {\bf p}_1 {\bf q}_1, m_2 m_3}^{ (0)}
\label{eq:6.1}
\end{eqnarray}
like the initial state
\begin{eqnarray}
\Phi_{0,a} \equiv \Phi_{ {\bf p}_1 {\bf q}_1, m_2 m_3}^{ 0} .
\label{eq:6.2}
\end{eqnarray}
Using the full Green's operator, $ G\equiv ( E + i \epsilon - H)^{ -1} $, 
to the full Hamiltonian,  $\Psi_{{\bf p}_1 {\bf q}_1, m_2 m_3}^{ (0)}$ 
obeys the equation
\begin{eqnarray}
    \Psi_{ {\bf p}_1 {\bf q}_1, m_2 m_3 }^{ (+)}  =   \Phi_{ {\bf p}_1 {\bf q}_1,m_2
m_3}^0 + G V \Phi_{ {\bf p}_1 {\bf q}_1,m_2 m_3}^0 \; .
\label{eq:6.3}
  \end{eqnarray}
A second scattering state is defined by
\begin{eqnarray}
\Psi_{ {\bf p}_1 {\bf q}_1, m_2 m_3 }^{ (-)}  =   \Phi_{ {\bf p}_1 {\bf q}_1,m_2 m_3}^0 +
G^* V \Phi_{ {\bf p}_1 {\bf q}_1,m_2 m_3}^0 \; .
\label{eq:6.4}
\end{eqnarray}
Both are related to each other as
\begin{eqnarray}
\Psi_{ {\bf p}_1 {\bf q}_1, m_2 m_3 }^{ (-)} & = &  \Psi_{ {\bf p}_1 {\bf q}_1, m_2 m_3
}^{ (+)} + ( G^* - G) V \Phi_{{\bf p}_1 {\bf q}_1,m_2 m_3}^0\cr
   & = &  \Psi_{ {\bf p}_1 {\bf q}_1, m_2 m_3 }^{ (+)} + 2 \pi i \delta( E - H) V
\Phi_{ {\bf p}_1 {\bf q}_1,m_2 m_3}^0 \; .
\label{eq:6.5}
   \end{eqnarray}
The S-matrix is defined as
 \begin{eqnarray}
 S_{ {\bf p}_1' {\bf q}_1', {\bf p}_1 {\bf q}_1}^{ m_2'm_3', m_2 m_2} \equiv 
\langle \Psi_{{\bf p}_1' {\bf q}_1', m_2' m_3'}^{ (-)} | 
\Psi_{ {\bf p}_1 {\bf q}_1, m_2 m_3}^{ (+)} \rangle .
\label{eq:6.6}
 \end{eqnarray}
Inserting Eq.~(\ref{eq:6.5}) leads to
\begin{eqnarray}
\lefteqn{S_{ {\bf p}_1' {\bf q}_1', {\bf p}_1 {\bf q}_1}^{ m_2'm_3', m_2 m_2} = 
  \langle \Psi_{ {\bf p}_1' {\bf q}_1', m_2' m_3'}^{ (+)}| 
\Psi_{ {\bf p}_1 {\bf q}_1, m_2 m_3}^{ (+)} \rangle } \cr
& &  - 2 \pi i \delta( E' - E) \langle \Phi_{ {\bf p}_1' {\bf q}_1',m_2'm_3'}^0| V
|\Psi_{ {\bf p}_1 {\bf q}_1 m_2 m_3}^{ (+)} \rangle.
\label{eq:6.7}
 \end{eqnarray}
Now, we have due to general considerations
\begin{eqnarray}
\langle   \Psi_{ {\bf p}_1' {\bf q}_1', m_2' m_3'}^{ (+)}| \Psi_{ {\bf p}_1 {\bf
q}_1, m_2 m_3}^{ (+)} \rangle = \langle \Phi_{ {\bf p}_1' {\bf q}_1',m_2' m_3'}^0
| \Phi_{ {\bf p}_1 {\bf q}_1,m_2 m_3}^0 \rangle.
\label{eq:6.8}
\end{eqnarray}
Consequently,
\begin{eqnarray}
S_{ {\bf p}_1' {\bf q}_1', {\bf p}_1 {\bf q}_1}^{ m_2'm_3', m_2 m_2} & = & 
  \langle \Phi_{ {\bf p}_1' {\bf q}_1',m_2' m_3'}^0| 
\Phi_{ {\bf p}_1 {\bf q}_1,m_2 m_3}^0 \rangle \cr
&  &-  2 \pi i \delta( E' - E) \langle \Phi_{ {\bf p}_1' {\bf q}_1',m_2' m_3'}^0| V
|\Psi_{ {\bf p}_1 {\bf q}_1, m_2 m_3}^{ (+)}\rangle \cr
  & = &  \langle \Phi_{ {\bf p}_1' {\bf q}_1',m_2'm_3'}^0| 
\Phi_{ {\bf p}_1 {\bf q}_1,m_2 m_3 }^0\rangle \cr
&   & -   2 \pi i \delta( E' - E) \langle \Phi_{ {\bf p}_1' {\bf q}_1',m_2' m_3'}^0| U^{
00} |\Phi_{ {\bf p}_1 {\bf q}_1,m_2 m_3}^0\rangle .
\label{eq:6.9}
 \end{eqnarray}
For the last equation we used the  definition of the transition operator $ U^{00}$.

Since the  scattering states in the definition of $S$ belong to the same Hamiltonian one has to have
\begin{eqnarray}
 S_{ {\bf p}_1' {\bf q}_1', {\bf p}_1 {\bf q}_1}^{ m_2'm_3', m_2 m_2} \equiv  \hat S_{
{\bf p}_1' {\bf q}_1', {\bf p}_1 {\bf q}_1}^{ m_2'm_3', m_2 m_2}
 \; \delta( E' - E) \; , 
\label{eq:6.10}
 \end{eqnarray}
 where $ E = \frac{ p_1^2}{m} + \frac{ q_1^2}{ 2 M_1}$ (setting the
$\alpha$-particle as spectator), and correspondingly as similar expression for $ E'$.
 The unitarity relation simply follows from the completeness relation spanned by the scattering states:
\begin{eqnarray}
\lefteqn{\langle   \Psi_{ {\bf p}_1' {\bf q}_1', m_2'm_3'}^{ (+)}| \Psi_{ {\bf p}_1
{\bf q}_1, m_2 m_3}^{ (+)}\rangle = }\cr
& &   \sum_{ m_2^{''} m_3^{''} } \int d^3 p_1^{''} d^3 q_1^{''}
\langle   \Psi_{ {\bf p}_1' {\bf q}_1', m_2'm_3'}^{ (+)}|  \Psi_{ {\bf p}_1^{''} {\bf
q}_1^{''}, m_2^{''}m_3^{''}}^{ (-)}\rangle \; 
 \langle \Psi_{ {\bf p}_1^{''} {\bf q}_1^{''}m_2^{''}m_3^{''}}^{ (-)}| \Psi_{
{\bf p}_1 {\bf q}_1, m_2 m_3}^{ (+)}\rangle,
\label{eq:6.11}
\end{eqnarray}
or in terms of the $S$- matrix elements
\begin{eqnarray}
\lefteqn{\langle \Phi_{ {\bf p}_1' {\bf q}_1',m_2'm_3'}^0| 
\Phi_{ {\bf p}_1 {\bf q}_1,m_2 m_3}^0\rangle  =} \cr
&  &  \sum_{ m_2^{''} m_3^{''} } \int d^3 p_1^{''} d^3 q_1^{''}
S_{{\bf p}_1{''} {\bf q}_1{''}, {\bf p}_1' {\bf q}_1'}^{ m_2{''}m_3{''}, m_2' m_3' *}
S_{{\bf p}_1{''} {\bf q}_1{''},{\bf p}_1 {\bf q}_1}^{ m_2{''}m_3{''}, m_2 m_3 }
\; .
\label{eq:6.12}
  \end{eqnarray}
This can be rewritten in terms of the matrix elements of $ U^{00}$. Using the
completeness relation, 
\begin{eqnarray}
\sum_{ m_2^{''} m_3^{''} } \int d^3 p_1^{''} d^3 q_1^{''} | \Phi_{ {\bf p}_1^{''}
{\bf q}_1^{''},m_2^{''}m_3^{''}}^0\rangle \; \langle 
\Phi_{ {\bf p}_1^{''} {\bf q}_1^{''},m_2^{''}m_3^{''}}^0| = 1 ,
\label{eq:6.13}
\end{eqnarray}
and
\begin{eqnarray}
\lefteqn{\delta( E^{ ''} - E') \langle \Phi_{ {\bf p}_1{''} {\bf q}_1{''},m_2^{''}m_3^{''}}^0|
U^{ 00} |\Phi_{ {\bf p}_1' {\bf q}_1', m_2'm_3'}^0\rangle = } \cr
 & & \langle \Phi_{ {\bf p}_1{''} {\bf q}_1{''},m_2^{''}m_3^{''}}^0| \delta( H_0 - E')  
U^{00} |\Phi_{ {\bf p}_1' {\bf q}_1', m_2'm_3'}^0\rangle ,
\label{eq:6.14}
\end{eqnarray}
leads to
\begin{eqnarray}
\lefteqn{\delta( E-E') \Big[ i \langle \Phi_{ {\bf p}_1 {\bf q}_1, m_2 m_3}^0| U^{ 00} 
|\Phi_{ {\bf p}_1' {\bf q}_1',m_2'm_3'}^0\rangle^* 
- i  \langle \Phi_{ {\bf p}_1' {\bf q}_1',m_2'm_3'}^0| U^{ 00} |\Phi_{ {\bf p}_1
  {\bf q}_1, m_2 m_3}^0\rangle } \cr
&  + &  2 \pi \sum_{ m_2^{''} m_3^{''} } \int d^3 p_1^{''} d^3 q_1^{''} \delta( E^{''} - E)\cr
& & \langle \Phi_{ {\bf p}_1^{''} {\bf q}_1^{''}, m_2^{''}m_3^{''}}^0| U^{ 00} |\Phi_{
{\bf p}_1' {\bf q}_1',m_2'm_3'}^0\rangle^*
\langle \Phi_{ {\bf p}_1^{''} {\bf q}_1^{''},m_2^{''}m_3^{''}}^0| U^{ 00} |\Phi_{
{\bf p}_1
{\bf q}_1,m_2 m_3}^0\rangle \Big]\cr
     &  = & 0 \; .
\label{eq:6.15}
\end{eqnarray}

More interesting is the partial wave decomposed version for the on shell matrix element
\begin{eqnarray}
\langle p_1'q_1' \alpha_1'| U^{00} | p_1 q_1 \alpha_1\rangle = U_{ \alpha_1' \alpha_1}^{ 00, J_1} ( p_1' p_1) \delta_{ J_1' J_1} \delta_{  M_1' M_1}
\label{eq:6.16}
\end{eqnarray}
     The dependence on $ p_1', p_1$ is sufficient since on shell
 $ q_1 = \sqrt{ 2 M_1 ( E - \frac{ p_1^2}{m})}, q_1'$ = $\sqrt{ 2 M_1 ( E' - \frac{ p_1^{'2}}{m})}$.

As shown in the Appendix~\ref{appendixc}, the matrix element 
 $  U_{ \alpha_1' \alpha_1}^{ 00, J_1} ( p_1' p_1)$ obeys the unitarity relation
\begin{eqnarray}
\lefteqn{ i \; U^{ 00, J *}_{ \alpha_1 \alpha_1'} ( p_1, p_1') - 
i \; U^{ 00,  J }_{ \alpha_1' \alpha_1} ( p_1', p_1) } \cr
&  + &  2 \pi  \sum_{ \alpha_1^{''}}\int  \delta( E^{''} - E)U^{ 00, J *}_{ \alpha_1^{''} \alpha_1'} ( p_1^{''}, p_1')
U^{ 00, J }_{ \alpha_1^{''} \alpha_1} ( p_1^{''}, p_1) = 0.
\label{eq:6.17}
   \end{eqnarray}
The corresponding relation for the partial wave projected S-matrix element is
\begin{eqnarray}
 \sum_{ \alpha_1^{''}}\int  \delta( E^{''} - E)\delta( E^{''} - E') \;
 S^{J*}_{ \alpha_1^{''}, \alpha_1'} ( p_1^{''} p_1') \; S^{J}_{ \alpha_1^{''},
\alpha_1} ( p_1^{''} p_1) 
  =   \delta_{ \alpha_1' \alpha_1} \frac{ \delta( p_1 -  p_1')}{ p_1^{2}}
\frac{ \delta( q_1 -  q_1')}{ q_1^{2}} .
\label{eq:6.18}
\end{eqnarray}
Note that not only discrete quantum numbers span the columns and 
rows of the $ S $-matrix but also the continuous quantum numbers
 $ p_1' p_1 $ which describe how the  energy is continuously distributed among the two relative motions.

\section{The capture process n+n+$\alpha \rightarrow {^6}$He}
\label{section7}

The matrix element for the capture process is simply related to the time 
reversed photo disintegration process of $^6$He into three free particles.
 It is well known~\cite{Golak:2005iy}  how to treat  photodisintegration  of $^3$He in
the Faddeev scheme. In essentially the same manner one can formulate
 photodisintegration of $^6$He based on an effective three-particle picture. 
Let $O$ be the
photon absorption  operator and $|\Psi_{^6\rm{He}}\rangle$ the $^6$He ground
state. The break up amplitude into $ nn\alpha$ can then be written as an infinite 
series of processes
\begin{eqnarray}
\lefteqn{\langle \Phi_{0,a} |U_0| \Psi_{^6\rm{He}}\rangle  = } \cr
& &  \langle \Phi_{0,a} | O |
\Psi_{^6\rm{He}}\rangle +
\sum_i \langle \Phi_{0,a} | V_i G_0 O | \Psi_{^6\rm{He}}\rangle \cr
 & +  &  \sum_{ij} \langle \Phi_{0,a} | V_i G_0 V_j G_0  O | \Psi_{^6\rm{He}}\rangle 
+ \cdots \; .
\label{eq:7.1}
\end{eqnarray}
Here $ V_i$ are the pair forces among the $nn$ and $n\alpha$ particles, and $
G_0$  is the free propagator. This infinite series in terms of pair forces represents 
final state interactions (FSI). The first term is the direct break up process 
generated by $O$.  Let us define
\begin{eqnarray}
\langle \Phi_{0,a} |U_0| \Psi_{^6\rm{He}}\rangle = \langle \Phi_{0,a} | O |
\Psi_{^6\rm{He}}\rangle + \sum_i
\langle \Phi_{0,a} |U_{ 0i}|  \Psi_{^6\rm{He}}\rangle ,
\label{eq:7.2}
\end{eqnarray}
where $ U_{0i}$ comprises all terms with $ V_i$ to the very left:
\begin{eqnarray}
U_{ 0i}|  \Psi_{^6\rm{He}}\rangle \equiv V_i G_0 O |  \Psi_{^6\rm{He}}\rangle + 
V_i \sum_j G_0 V_j\; G_0\; O |  \Psi_{^6\rm{He}}\rangle + \cdots  \; .
\label{eq:7.3}
\end{eqnarray}
Clearly this can be summed up as
\begin{eqnarray}
U_{ 0i}|  \Psi_{^6\rm{He}}\rangle =  V_i \; G_0 \; O |\Psi_{^6\rm{He}}\rangle +
V_i\; G_0 \sum_j U_{0j} | \Psi_{^6\rm{He}}\rangle .
\label{eq:7.4}
 \end{eqnarray}
Separating the terms $ U_{ 0i}| \Psi_{^6\rm{He}}\rangle$ to the left 
and introducing the $t$-matrices
$ t_i$ leads to  three coupled Faddeev equations ($ i= 1,2,3$),
\begin{eqnarray}
U_{ 0i}|  \Psi_{^6\rm{He}}\rangle =  t_i G_0 O |  \Psi_{^6\rm{He}}\rangle + t_i G_0 \sum_{ j \ne i}
U_{0j} |  \Psi_{^6\rm{He}}\rangle .
\label{eq:7.5}
\end{eqnarray}

  The photon absorption operator $ O$ has to be symmetric under the exchange of
the two neutrons, which we number as particles $2$ and $3$. Thus, 
using the antisymmetry of $|  \Psi_{^6\rm{He}}\rangle$ with respect to the two
neutrons, one finds 
  \begin{eqnarray}
  P_{23} U_{ 02}|  \Psi_{^6\rm{He}}\rangle = - U_{ 03}|  \Psi_{^6\rm{He}}\rangle.
\label{eq:7.6}
  \end{eqnarray}
This leads to the  two coupled equations
  \begin{eqnarray}
U_{ 01}|  \Psi_{^6\rm{He}}\rangle & = &   t_1 G_0 O |  \Psi_{^6\rm{He}}\rangle + t_1 G_0 ( 1 -
P_{23}  U_{02}) |  \Psi_{^6\rm{He}}\rangle \cr
U_{ 02}|  \Psi_{^6\rm{He}}\rangle  & = &   t_2 G_0 O |  \Psi_{^6\rm{He}}\rangle  + t_2 G_0 ( U_{ 01}
- P_{23}  U_{02}) |  \Psi_{^6\rm{He}}\rangle ,
\label{eq:7.7}
\end{eqnarray}
corresponding to Eqs.~(\ref{eq:1.12}) from Sect.~\ref{section2}. 

The complete break up amplitude is given by
\begin{eqnarray}
\langle \Phi_{0,a} |U_0| \Psi_{^6\rm{He}}\rangle  & = &  \langle \Phi_{0,a} | O |
\Psi_{^6\rm{He}}\rangle  +
\langle \Phi_{0,a} |U_{ 01}|  \Psi_{^6\rm{He}}\rangle \cr
&  + &  \langle \Phi_{0,a} | ( 1 - P_{23}) U_{ 02}| \Psi_{^6\rm{He}}\rangle .
\label{eq:7.8}
\end{eqnarray}
Using adequate pair forces and photon absorption operators 
(single particle currents, two-body currents and possibly beyond) these  coupled 
equations can be solved by standard techniques \cite{Golak:2005iy}. 
%That capture process is of astrophysical relevance \cite{efros}, \cite{goerres}.

\section{ Summary}

The structure inherent in the continuum states of the n+n+$\alpha$ system has  so far
only been explored in the framework of the HH
approach~\cite{danilin3,Fedorov:2003jx,Danilin:2004mx,Danilin:2006qq,Danilin:2007zz,ershov3}.
There are strong initial and final state interaction peaks, not only in the  $nn$ 
subsystem but also in the $n-\alpha$ subsystem. This poses a still unsolved challenge
for the expansion into the discrete set of K-harmonics as already known for the 
$ n+d \rightarrow n+n+p$ system. This is pointed out by the authors of
Ref.~\cite{danilin3}, who note, that even for a maximum $K_{max}=20$ in their
calculation, the result is not completely converged.

A corresponding investigation  in the Faddeev approach is still missing. The aim of 
this paper is to lay the formal ground to do so. 

In the Faddeev approach all the structures in the relative motions of the three
particles are mapped out correctly, thus leading to 
a reliable path to the three-to-three scattering S-matrix, 
which contains the information  of the resonance structure of the $^6$He system. 

We derived two
 coupled Faddeev equations for the three-to-three scattering amplitudes. In a 
partial wave decomposed representation they form a system of two-dimensional 
coupled equations for each fixed total angular momentum. 
The multiple scattering
series being arranged in  powers in the two-body $t$-matrices is the natural starting
point for the solution of this coupled system of integral equations. 
The term linear in the $t$-matrices is
disconnected. The next term, second order in $t$, has well established logarithmic
singularities in the external  momenta. Only the term  of third order in $t$ is a smooth
function of the external momenta and thus can serve as  driving term for the
consecutive application  of the Faddeev kernels. This provides all higher order
terms which can then summed up by Pad\'e.

Since up to now nearly all Faddeev based investigations  of the discrete structure of
 the n+n+$\alpha$ system are based on finite rank forces we also 
derived the continuum equations using this type of forces.
The unitarity relations are especially interesting since the rows and columns of 
the S- matrix are  not only numbered by discrete quantum numbers but also 
by continuously varying  on-shell momenta.  

Finally we provided  Faddeev equations for
the  n+n+$\alpha$ capture process to the
 $^6$He  ground state.
We pointed out that it is not necessary here to first evaluate the
three-to-three wave function as has been done in Ref.~\cite{vanMeijgaard:1992zz},
 and that one directly can use a Faddeev form
for the entire break-up amplitude with no disconnected  terms as is done in modern
calculations
~\cite {Golak:2005iy, wgphysrep}, and as it was pioneered in ~\cite{barbour,gibson}.

This capture process is relevant for the production rate
of $^6$He  in astrophysical environments~\cite{deDiego:2010sf}
characterized by high neutron and alpha densities, e.g. those related to supernova shock
fronts. In  Ref.~\cite{efros} this three-body process is approximated by sequential
two-body processes, whereas in principle a genuine three-body reactions needs to
be calculated.  
Very recently it has been pointed out~\cite{deDiego:2010sf} that  currently employed
two-step mechanisms over intermediate resonances in the three-to-three scattering of
the $n+n+\alpha$ system are most likely insufficient, since the time delays for those
intermediate steps are comparable to the duration of the entire process. This strongly
supports the need for the approach we present in this paper.

%---------------------------------------------------------------------------

\vfill

\section*{Acknowledgments}
This work was performed in part under the
auspices of the U.~S.  Department of Energy under contract
No. DE-FG02-93ER40756 with Ohio University and grant 163.PNE.A912371 from the
Minist\`ere de L'enseignement Superieur et de la Recherche Scientifique, Algeria (K.K).
We (K.K and Ch.E) thank the Institute for Nuclear Theory at the University of Washington
for its hospitality and the U.~S.  Department of Energy for partial support during the
initial phase of this work.

%-------------------------------------------------------------------------------
%****************************************************************************

%\newpage
\appendix

\section{Partial Wave Decomposition in the $nn\alpha$ System}
\label{appendixa}

  Partial wave decomposition of three-body wave functions have often been documented, 
see for instance \cite{wgphysrep} and focus on the $ nn \alpha$- system~\cite{khalida}. 
Thus we are here relatively brief.

The projection of the free state, Eq.~(\ref{eq:1.13})  
onto the partial wave basis states of Eq.~(\ref{eq:1.17}) is given by
\begin{eqnarray}
\lefteqn{\langle p_1' q_1' \alpha_1'| \Phi_{{\bf p}_1 {\bf q}_1 m_2 m_3}\rangle =} \cr
 & &  \left( 1 + (-)^{ l_1' + s_1'}\right) \sum_\mu 
( j_1' \lambda_1' J', \mu,  M' - \mu)  ( l_1' s_1' j_1', \mu - m_2 -m_3, m_2 + m_3)\cr
 & & \left( \frac{1}{2}\frac{1}{2} s_1', m_2 m_3 \right) \; 
\frac{\delta( p_1' -p_1)}{ p_1^2} \frac{ \delta(q_1' -q_1)}{ q_1^2} \;
 Y_{ l_1'\mu - m_2 -m_3} ^* ( \hat p_1)  \; Y_{ \lambda_1' M' - \mu}^* (\hat q_1) .
\label{eq:a.1}
\end{eqnarray}
This leads to
\begin{eqnarray}
\lefteqn{\langle p_1^{''} q_1^{''} \alpha_1^{''}| t_1 |\Phi_{{\bf p}_1 {\bf q}_1 m_2
m_3}\rangle =} \cr
& &  \int dp_1' p_1^{'2} t_{ \alpha_1^{''}} ( p_1^{''} p_1' q_1^{''} ) \;
\Bigg[ \left( 1 + (-)^{ l_1^{''} + s_1^{''}}\right) \sum_{\mu} ( j_1^{''} \lambda_1^{''} J', \mu,  M^{''} - \mu) \cr
 & & ( l_1^{''} s_1^{''} j_1^{''}, \mu - m_2 -m_3, m_2 + m_3) \left(
\frac{1}{2}\frac{1}{2} s_1^{''}, m_2 m_3 \right) \; \frac{ \delta( p_1' -p_1)}{ p_1^2}
  \frac{ \delta( q_1^{''} -q_1)}{ q_1^2} \cr
 & &  Y_{ l_1^{''} \mu - m_2 -m_3} ^* ( \hat p_1) Y_{ \lambda_1^{''} M^{''} - \mu} ^* ( \hat q_1)]\cr
  & & \equiv \frac{ \delta( q_1^{''} -q_1)}{ q_1^2} \; t_{ \alpha_1^{''}} (
p_1^{''} p_1, E_{q_1}) \; C_{\alpha_1^{''}} ^{ m_2 + m_3}( \hat p_1 \hat q_1),
 \end{eqnarray}
and gives  the driving term of Eq.~(\ref{eq:1.18}) together with the amplitude 
$C$ given in Eq.~(\ref{eq:1.19}).

\noindent
In case of the second driving term in Eq.~(\ref{eq:1.14}) it is  adequate to 
rewrite the free state $
|\Phi_{{\bf p}_1 {\bf q}_1 m_2 m_3}\rangle $ in terms of the Jacobi momenta
 of the type 2 (where the neutron is the spectator):
 \begin{eqnarray}
 {\bf p}_2 & = &  - \beta {\bf p}_1 - \gamma {\bf q}_1\cr
 {\bf q}_2 & = &  {\bf p}_1 - \alpha {\bf q}_1 \; ,
 \end{eqnarray}
with
\begin{eqnarray}
\alpha & = &  \frac{1}{2}\cr
\beta & = &  \frac{ m_{\alpha}}{ m + m_{\alpha}}\cr
\gamma & = &  \frac{ 2 m + m_{\alpha}}{ 2( m + m_{\alpha})}\; ,
\end{eqnarray}
 and $ \alpha \beta + \gamma =1 $.

\noindent
Then,   
\begin{eqnarray}
\lefteqn{|\Phi_{{\bf p}_1 {\bf q}_1 m_2 m_3}\rangle =} \cr
& & | {\bf p}_2 {\bf q}_2\rangle | 0\rangle_1
| m_2\rangle_2 | m_3\rangle_3 \Big| 0\left( \frac{1}{2}\frac{1}{2}\right) 1 \Big\rangle
  -  | \tilde {{\bf p}}_2 \tilde {{\bf q}}_2 \rangle | 0\rangle_1 | m_2\rangle_3 |
     m_3\rangle_2 \cr
  & & | 0( \frac{1}{2}\frac{1}{2}) 1\rangle \; ,
\end{eqnarray}
where
\begin{eqnarray}
 \tilde {{\bf p}}_2 & = &   \beta {\bf p}_1 - \gamma {\bf q}_1\cr
 \tilde {{\bf q}}_2 & = & -  {\bf p}_1 - \alpha {\bf q}_1 \; .
 \end{eqnarray}
The partial wave projected state in system ``2'' is given by
  \begin{eqnarray}
\lefteqn{\langle p_2' q_2' \alpha_2'|\Phi_{{\bf p}_1 {\bf q}_1 m_2 m_3}\rangle =
} \cr
& & \frac{\delta( p_2' - p_2)}{ p_2^2} \frac{ \delta( q_2' - q_2)}{ q_2^2}
\; \delta_{ s_2' \frac{1}{2}} \cr
& & \sum_{\mu} ( j_2' I_2' J', \mu M'-\mu)( l_2'\frac{1}{2}  j_2', \mu - m_3, m_3)   Y_{ l_2' \mu - m_3}^* ( \hat p_2)\cr
& & ( \lambda_2'  \frac{1}{2} I_2', M' -\mu - m_2, m_2) Y_{ \lambda_2'M' -\mu - m_2}^* ( \hat q_2)\cr
& - & \frac{ \delta( p_2' - \tilde p_2)}{ {\tilde p_2}^2}
\frac{ \delta( q_2' - \tilde q_2)}{ {\tilde q_2}^2} \; \delta_{ s_2' \frac{1}{2}} \cr
& & \sum_{\mu} ( j_2' I_2' J', \mu M'-\mu)( l_2'\frac{1}{2}  j_2', \mu - m_2, m_2)   Y_{
l_2' \mu - m_2}^* ({\hat {\tilde p}}_2)\cr
& & ( \lambda_2'  \frac{1}{2} I_2', M' -\mu - m_3, m_3) Y_{ \lambda_2'M' -\mu - m_3}^*
({\hat {\tilde q}}_2) \; ,
\end{eqnarray}
and the second driving term becomes
\begin{eqnarray}
\lefteqn{ \langle p_2' q_2' \alpha_2'| t_2 |\Phi_{{\bf p}_1 {\bf q}_1 m_2
m_3}\rangle \equiv } \cr
& &\frac{ \delta( q_2' - q_2)}{ q_2^2} \; t_{ \alpha_2'} ( p_2' p_2, E_{ q_2'})
\; D_{ \alpha_2'}^{ m_2, m_3} ( \theta_{p_2} \theta_{q_2})\cr
  & - & \frac{ \delta( q_2' - \tilde q_2)}{ {\tilde q_2}^2} \; 
t_{ \alpha_2'} ( p_2' \tilde p_2, E_{ \tilde q_2}) \;
  \tilde D_{ \alpha_2'}^{ m_2, m_3} ( \theta_{\tilde p_2} \theta_{\tilde q_2}) ,
 \end{eqnarray}
with $D$ and $ \tilde D$ given in Eqs.~(\ref{eq:1.20}) and (\ref{eq:1.21}). 
For the kernel pieces we refer to \cite{khalida}.

%%%%%%%%%%%%%%%%%%%%%%%%%%%%%%%%%%%%%%%%%%%%%%%%%%%%%%%%%%%%%%

\section{Avoiding logarithmic singularities in the integrals} 
\label{appendixb}

We illustrate the new manner to rewrite the Faddeev kernel such that only a single 
pole singularity appears in an example (for more details
see~\cite{Witala:2008my,Elster:2008hn})
Consider the first kernel in Eq.~(\ref{eq:1.18}), of which the first piece can be
rewritten as
\begin{eqnarray}
\lefteqn{ \int dx \int dq_2' q_2^{'2} \sum_{\alpha_2'} \int dp_1^{''} p_1^{''2} \;
 \frac{ \delta( p_1^{''} - \pi_1 ( q_1' q_2' x))}{ p_1^{''2}} \;
 t_{ \alpha_1'} ( p_1' p_1^{''}, E_{q_1'}) \;  G_0( p_1^{''}, q_1') } \cr
  & & G_{ \alpha_1' \alpha_2'} ( q_1' q_2' x) \int dp_2' p_2^{'2} \; \frac{ \delta( p_2' -
\pi_2 ( q_1' q_2' x))}{ p_2^{'2}} \;  \langle p_2' q_2' \alpha_2'| U_{2,a} \; .
\label{eq:b.1}
  \end{eqnarray}
The two $\delta$-functions are then changed according to
\begin{eqnarray}
\delta\big( p_1^{''}- \pi_1( q_1' q_2' x)\big) &=&  \frac{ 2 p_1^{''}}{ 2 \alpha q_1'
q_2'} \; \delta( x - x_0) \; \Theta( 1 -| x_0|) \cr
\delta \big( p_2'- \pi_2( q_1' q_2' x) \big) &=&  \delta \left( p_2' - \sqrt{ \gamma
q_1^{'2} + \frac{ \beta}{\alpha} p_1^{''2} - \frac{ \beta \gamma}{ \alpha} q_2^{'2}}
\right) \cr
& & \Theta \left( \gamma q_1^{'2} + \frac{ \beta}{\alpha} p_1^{''2} - \frac{ \beta
\gamma}{ \alpha} q_2^{'2} \right) \; ,
\label{eq:b.2}
\end{eqnarray}
with
\begin{eqnarray}
x_0 & = &  \frac{ p_1^{''2} - \alpha^2 q_1^{'2} - q_2^{'2}} { 2 \alpha q_1' q_2'} \; .
\label{eq:b.3}
\end{eqnarray}
Inserting this into Eq.~(\ref{eq:b.1}) leads to
\begin{eqnarray}
& &   \int dx \int dq_2' q_2^{'2} \sum_{\alpha_2'} \int dp_1^{''} p_1^{''2} \;
   \frac{ 2 p_1^{''}}{ 2 \alpha q_1' q_2'}  \;\delta( x - x_0)  \;\Theta( 1 -| x_0|)  \;
 t_{ \alpha_1'} ( p_1' p_1^{''}, E_{q_1'}) \cr
& &  G_0( p_1^{''}, q_1')
 \; G_{ \alpha_1' \alpha_2'} ( q_1' q_2' x)  \int dp_2' p_2^{'2} \;
  \delta \left( p_2' - \sqrt{ \gamma q_1^{'2} + \frac{ \beta}{\alpha} p_1^{'2} - \frac{
\beta \gamma}{ \alpha} q_2^{'2}} \right) \cr
& & \Theta \left( \gamma q_1^{'2} + \frac{ \beta}{\alpha} p_1^{'2} - \frac{ \beta \gamma}{
\alpha} q_2^{'2} \right) \langle p_2' q_2' \alpha_2'| U_{2,a}\cr
& = & \frac{1}{\alpha q_1'} \int dq_2' q_2' \sum_{\alpha_2'} \int dp_1^{''} p_1^{''} \; 
t_{ \alpha_1} ( p_1' p_1^{''}, E_{q_1'}) \;  G_0( p_1^{''}, q_1')\cr
& & G_{ \alpha_1' \alpha_2'} ( q_1' q_2' x_0) \Big\langle \sqrt{ \gamma q_1^{'2} + 
\frac{ \beta}{\alpha} p_1^{'2} - \frac{ \beta \gamma}{ \alpha} q_2^{'2}} \;  q_2'
\alpha_2'\Big| U_{2,a} \cr
& & \Theta \left( 1 -\Big|\frac{ p_1^{''2} - \alpha^2 q_1^{'2} - q_2^{'2}} { 2 \alpha q_1'
q_2'}  \Big| \right) \; 
 \Theta \left( \gamma q_1^{'2} + \frac{ \beta}{\alpha} p_1^{''2} - \frac{ \beta \gamma}{
\alpha} q_2^{'2} \right) \; .
\label{eq:b.4}
\end{eqnarray}
The two $\Theta$-functions define the domain D for the integrations over 
$ p_1^{''}$ and $ q_2'$. Thus we end up with
\begin{eqnarray}
\lefteqn{\frac{1}{\alpha q_1'}\int dp_1^{''} p_1^{''} \; 
t_{ \alpha_1} ( p_1' p_1^{''}, E_{q_1'})
\; \frac{ 1}{ E + i \epsilon - \frac{p_1^{''2}}{m} - \frac{ q_1^{'2}}{ 2 M_1}} } \cr
& &\int_{| p_1^{''} - \alpha q_1'|}^{p_1^{''} + \alpha q_1'} dq_2' q_2' \sum_{\alpha_2'}
  G_{ \alpha_1' \alpha_2'} ( q_1' q_2' x_0) \Big\langle \sqrt{ \gamma q_1^{'2} + \frac{
\beta}{\alpha} p_1^{'2} - \frac{ \beta \gamma}{ \alpha} q_2^{'2}} \; 
 q_2' \alpha_2' \Big| U_{2,a} \; .
\label{eq:b.5}
\end{eqnarray}

\noindent
The singularity in $ G_0( p_1^{''}, q_1')$   is now a single pole in $ p_1^{''}$ for a 
 given $ q_1'$. This type of singularity does not pose any numerical problem and can be
implemented with standard techniques~\cite{RHLandau}. Note that for 
$ q_1' \ge \sqrt{ 2M_1 E} $ there is no pole and one might as well keep the original form.

%%%%%%%%%%%%%%%%%%%%%%%%%%%%%%%%%%%%%%%%%%%%%%%%%%%%%%%%%%%%%%

\section{Partial Wave Decomposed Transition Amplitude}
\label{appendixc}

The definition of the partial wave decomposed transition amplitude is
\begin{eqnarray}
\lefteqn{ \langle \Phi_{{\bf p}_1' {\bf q}_1' m_2' m_3'}| U^{00} | 
\Phi_{{\bf p}_1 {\bf q}_1 m_2 m_3}\rangle  \equiv } \cr
& & \sum_{ \tilde {\alpha_1}} \int d \tilde {p_1} 
{\tilde {p_1}}^2 d \tilde {q_1}  {\tilde {q_1}}^2
   \langle \Phi_{{\bf p}_1' {\bf q}_1' m_2' m_3'}| \tilde {p_1}' \tilde {q_1}' \tilde
{\alpha_1}'\rangle  \cr
& &  \langle  \tilde {p_1}' \tilde {q_1}' \tilde {\alpha_1}'|  U^{00}|
    \tilde {p_1} \tilde {q_1} \tilde {\alpha_1} \rangle \langle \tilde {p_1} \tilde
{q_1} \tilde {\alpha_1} |\Phi_{{\bf p}_1 {\bf q}_1 m_2 m_3}\rangle \; .
\label{eq:c.1} 
\end{eqnarray}
This inserted into Eq.~(\ref{eq:6.15}) yields
\begin{eqnarray}
\lefteqn{\delta( E-E')   \sum_{ \tilde {\alpha_1}'}\int \sum_{ \tilde {\alpha_1}}\int
 \langle \tilde {p_1} \tilde {q_1} \tilde {\alpha_1}| \Phi_{ {\bf p}_1 {\bf q}_1, m_2
m_3}\rangle \langle \Phi_{ {\bf p}_1' {\bf q}_1',m_2'm_3'| \tilde {p_1}' \tilde  {q_1}'
 \tilde {\alpha_1}'}\rangle } \cr
 & & \Bigg[ i \; \langle \tilde {p_1} \tilde {q_1} \tilde {\alpha_1} | U^{ 00}| \tilde {p_1}'
\tilde {q_1}' \tilde {\alpha_1}'\rangle^*
  -i \langle \tilde {p_1}' \tilde {q_1}' \tilde {\alpha_1}'| U^{ 00}| \tilde {p_1}
\tilde {q_1} \tilde {\alpha_1} \rangle \cr
&+&  2 \pi  \sum_{ m_2^{''} m_3^{''}} \int d^3 p_1^{''}d^3 q_1^{''} \; 
\delta( E^{''} - E)  \cr
& & \sum_{ \tilde {\alpha_1}^{''}}
 \int \langle \Phi^{''} |  \tilde  {p_1}^{''} \tilde {q_1}^{''} \tilde
{\alpha_1}^{''}\rangle^*
  \langle  \tilde {p_1}^{''} \tilde {q_1}^{''} \tilde {\alpha_1}^{''}| U^{ 00}|\tilde
{p_1}' \tilde {q_1}' \tilde {\alpha_1}'\rangle^* \cr
& & \sum_{ \tilde {\alpha_1}^{'''}}\int \langle \Phi^{''} |  \tilde  {p_1}^{'''} \tilde
{q_1}^{'''}\tilde {\alpha_1}^{'''}\rangle
 \langle \tilde {p_1}^{'''} \tilde {q_1}^{'''}\tilde {\alpha_1}^{'''}| U^{ 00}| \tilde
{p_1} \tilde {q_1} {\tilde \alpha_1}\rangle \Bigg] \cr
 & = & 0 \; ,
\label{eq:c.2}
 \end{eqnarray}
which, using the completeness relation,
\begin{eqnarray}
\lefteqn{\sum_{ m_2^{''} m_3^{''}}\int d^3 p_1^{''}d^3 q_1^{''} \delta( E^{''} - E)
\langle \Phi^{''} |  \tilde {p_1}^{''} \tilde {q_1}^{''} \tilde
{\alpha_1}^{''}\rangle^* \langle \Phi^{''} |  \tilde  {p_1}^{'''} \tilde {q_1}^{'''}
\tilde {\alpha_1}^{'''}\rangle = } \cr
&  &  \frac{ \delta( \tilde {p_1}^{''} - \tilde {p_1}^{'''})}{ (\tilde {p_1}^{''})^2}
\frac{ \delta( \tilde {q_1}^{''} - \tilde {q_1}^{'''})}{ (\tilde {q_1}^{''})^2}\delta_{
\tilde {\alpha_1}^{''} \tilde {\alpha_1}^{'''}}  \; \delta( E^{''} - E) \; ,
\label{eq:c.3}
\end{eqnarray}
leads to
\begin{eqnarray}
\lefteqn{ \delta( E-E')   \sum_{ \tilde {\alpha_1}'}\int \sum_{ \tilde {\alpha_1}}\int
 \langle \tilde {p_1} \tilde {q_1} \tilde {\alpha_1} | \Phi_{ {\bf p}_1 {\bf q}_1, m_2
m_3} \rangle \langle \Phi_{ {\bf p}_1' {\bf q}_1',m_2'm_3'}| \tilde {p_1}' \tilde {q_1}'
 \tilde {\alpha_1}'\rangle } \cr
 & & \Bigg[ i   \langle \tilde {p_1} \tilde {q_1} \tilde {\alpha_1}|U^{ 00}| \tilde {p_1}'
\tilde  {q_1}' \tilde {\alpha_1}'\rangle^*
  -i \langle \tilde  {p_1}' \tilde {q_1}'\tilde {\alpha_1}'|U^{ 00}| \tilde {p_1} \tilde
{q_1} \tilde {\alpha_1} \rangle \cr
&  + &  2 \pi  \sum_{ \tilde {\alpha_1}^{''}}\int  \delta( E^{''} - E)
   \langle  \tilde {p_1}^{''} \tilde {q_1}^{''} \tilde {\alpha_1}^{''}| U^{ 00}| \tilde
{p_1}' \tilde {q_1}' \tilde {\alpha_1}'\rangle^* \cr
& &  \langle \tilde {p_1}^{''} \tilde {q_1}^{''}\tilde {\alpha_1}^{''}| U^{ 00}| \tilde
{p_1} \tilde {q_1}  \tilde {\alpha_1}\rangle \Bigg] \cr
 & = & 0 \; .
\label{eq:c.4}
 \end{eqnarray}

\noindent
Then using Eqs.~(\ref{eq:a.1}) and (\ref{eq:6.16}), the orthogonality of the 
spherical harmonics and of the Clebsch-Gordan coefficients, 
one can project onto the on s-shell unitarity
 relation of Eq.~(\ref{eq:6.17}).

%%%%%%%%%%%%%%%%%%%%%%%%%%%%%%%%%%%%%%%%%%%%%%%%%%%%%%%%%%%%%%%%%%%%%%%%%%

%%%%%%%%%%%%%%%%%%%%%%%%%%%%%%%%%%%%%%%%%%%%%%%%%%%%%%

%\clearpage

\vfill

\end{document}